\documentstyle[prd,aps,epsfig,eqsecnum,amssymb,amsmath,floats]{revtex}
\begin{document}
\title{
Constraints on diffuse neutrino background from primordial 
black holes
}
\author{
E. V. Bugaev and K. V. Konishchev
}
\address{
Institute for Nuclear Research, Russian Academy of Sciences, 
Moscow 117312, Russia
}
\date{
\today
}
\maketitle

\begin{abstract}
We calculated the energy spectra and the fluxes of electron neutrinos 
in extragalactic space emitted
in the process of the evaporation of primordial black holes (PBHs) in the early
universe. It was assumed that PBHs are formed by a blue power-law spectrum
of primordial density fluctuations.
In the calculations of neutrino spectra the spectral index of density 
fluctuations and the reheating temperature were used as free parameters. 
The absorption of
neutrinos during propagation in the space was taken into account.
We obtained the bounds on the spectral index
assuming validity of the standard picture of
gravitational collapse and using the available data of several experiments
with atmospheric and solar neutrinos. The comparison of our results with the
previous constraints (which had been obtained using diffuse photon background
data) shows that such bounds are quite sensitive to an assumed form of the
initial PBH mass function.
\end{abstract}

\section{Introduction}
\label{sec:int}
Some recent inflation models (e.g., the hybrid inflationary scenario \cite{1})
predict the "blue" power-spectrum of primordial density fluctuations. In turn,
as is well known, the significant abundance of primordial black holes (PBHs) 
is possible just in the case when the density fluctuations have an $n>1$ 
spectrum ($n$ is the spectral index of the initial density fluctuations, 
$n>1$ spectrum is, by definition, the "blue perturbation spectrum").
 
   Particle emission from PBHs due to the evaporation process predicted by
Hawking \cite{2} may lead to observable effects. Up to now, PBHs have not been 
detected, so the observations have set limits on the initial PBH abundance
or on characteristics of a spectrum of the primordial density fluctuations.
In particular, PBH evaporations contribute to the extragalactic neutrino 
background. The constraints on an intensity of this background (and, 
correspondingly, on an PBH abundance) can be obtained from the existing 
experiments
with atmospheric and solar neutrinos. The obtaining of such constraints is
a main task of the present paper.

 The spectrum and the intensity of the evaporated neutrinos depend heavily on 
the PBH's mass. Therefore, the great attention should be paid to the 
calculation
of the initial mass spectrum of PBHs. We use in this paper the following 
assumptions leading to a prediction of the PBH's mass spectrum.

1. The formation of PBHs begins only after an inflation phase when the universe
returns to the ordinary radiation-dominated era. The reheating process is such
that an equation of state of the universe changes almost instantaneously 
into the radiation type (e.g., due to the parametric resonance \cite{3}) after
the inflation.

2. It is assumed, in accordance with analytic calculations \cite{4,5} that a 
critical 
size of the density contrast needed for the PBH formation, $\delta_c$, is 
about $1/3$. Further, it is assumed that all PBHs have mass roughly equal
to the horizon mass at a moment of the formation, independently of the 
perturbation size.

3. Summation over all epochs of the PBH formation can be done using the
Press-Schechter formalism \cite{6}. This formalism is widely used in the 
standard
hierarchial model of the structure formation for calculations of the mass
distribution functions (see, e.g., \cite{7}).

It was shown recently that near the threshold of a black hole formation the 
gravitational collapse
behaves as a critical phenomenon \cite{8}. In this case the initial mass 
function will be quite 
different from that which follows from the standard calculations of 
refs.~\cite{5,7}.The first calculations
of the PBH initial mass function (e.g., in ref.\cite{8}) have been done 
under the assumption that 
all PBHs form at the same horizon mass (by other words, that all PBHs form 
at the smallest horizon 
scale immediately after reheating). The initial PBH mass spectrum for the 
case of the critical
collapse, based on the Press-Schechter formalism, was obtained in the 
recent work \cite{9}.

The calculations in the present paper are based on the standard 
\cite{4,5} picture of the 
gravitational collapse leading to a PBH formation. The case of the 
critical collapse will
be considered in a separate work.

The plan of the paper is as follows.

In Sec.\ref{sec:PBHsp} we give, for completeness, the brief derivation 
of a general formula
for the initial PBH mass spectrum. The final expression is presented 
in a form, which is
valid for an arbitrary relation between three physical values:
the initial PBH mass, the fluctuation mass (i.e., the mass of the 
perturbed region) at a moment of 
the collapse, and the density contrast in the perturbed region 
(also at a moment of the collapse).
This expression contains the corresponding results obtained in 
refs.~\cite{7,9} as particular cases.
As in refs.~\cite{7,9}, the derivation is based on the linear perturbation 
theory and on the assumption 
that a power spectrum of the primordial fluctuations can be described by 
a power law.

In Sec.\ref{sec:neutbackgr} we derive the approximate formula for a 
calculation of the extragalactic
neutrino background from PBH evaporations. We do not use cosmological 
models of the inflation and of the spectrum
of primordial fluctuations, so there are two free parameters: the reheating 
temperature and the spectral index.
At the end of the section, some examples of instantaneous neutrino spectra 
from the evaporation of an 
individual black hole are presented.

In Sec.\ref{sec:normalization} we perform the normalization of the primordial power spectrum
of density perturbations by COBE data on the angular power spectrum of CMB temperature 
fluctuations. We consider the case of flat cosmological models with nonzero cosmological
constant. We show, that the result of this normalization rather weakly depends on the matter
content of the present Universe, $\Omega_m=1-\Omega_{\Lambda}$.

In Sec.\ref{sec:spectra} we present the results of numerical calculations of the neutrino 
background spectra from PBH evaporations, with taking into account the effects of neutrino
absorption in the space. For background neutrino energies
$\sim 10-100\;MeV$ we studied relative contributions to the background intensity of 
different cosmological redshifts (and it is shown that at high reheating temperatures 
the characteristic values of the redshifts are very large).

In Sec.\ref{sec:constr} we consider the possibilities of constraining the spectral index,
using experiments with neutrinos of natural origin.

In Sec.\ref{sec:res} the spectral index constraints, followed from our calculations and from available data
of the neutrino experiments, are given.

\section{The initial mass spectrum of PBHs}
\label{sec:PBHsp}
As is pointed out in the {\bf Introduction}, we use the Press-Schechter formalism which allows to carry out the 
summation over all epochs of PBHs formation. According to this formalism,
the mass spectrum of density fluctuations ( i.e., the number density of regions with mass 
between $M$ and $M+dM$ ) is calculated by the formula
\begin{equation}
\label{1}
n\left( M,\delta _c \right)dM=\frac{\rho _i }{M}
\left| \frac{\partial \beta}{\partial M}(M,\delta _c )\right| dM\;\;\;.
\end{equation}
Here, $\beta (M,\delta _c)$ is the fraction of regions having sizes 
larger than $R$ and density contrast larger than $\delta _c$ ,
\begin{equation}
\label{2}
\beta(M,\delta _c)=2\int^\infty_{\delta _c} P(M,\delta)d\delta\;\;\;,
\end{equation}
\begin{equation}
\label{3}
P(M,\delta)=\frac{1}{\sqrt{2\pi}\sigma _R (M)}\exp\left(-\frac{\delta^2}
{2\sigma _R^2 (M)} \right)\;\;\;.
\end{equation}
In these formulas, $\delta$ is the initial density contrast, $\sigma_R$ is the standard
deviation of the density contrast of the regions with a comoving size $R$ and mass $M$.
The factor of $2$ in Eq.(\ref{2}) ensures the correct normalization of 
$n(M,\delta_c)$, integrated over all $M$. This factor is usually added to take
into account (approximately) the contribution of underdense regions to
gravitationally bound objects \cite{7}. 

It is convenient to introduce the double differential distribution 
$n(M,\delta)$, the integral over which gives the total number of fluctuated
regions:
\begin{equation}
n=\int\!\!\!\!\int n(M,\delta)dMd\delta \;\;\;\;,\;\;\;
n(M,\delta_c)=\int n(M,\delta)d\delta\;\;\;.
\end{equation}
Using Eqs.(\ref{1}-\ref{3}), for $n(M,\delta)$ one has the expression
\begin{equation}
\label{25}
n(M,\delta)=\sqrt{\frac{2}{\pi}}\frac{\rho_i}{M}\frac{1}{\sigma_R^2 (M)}\frac{\partial \sigma_R}{\partial M}
\left|\left(\frac{\delta^2}{\sigma_R^2 (M)}-1\right)\right|\exp\left(
-\frac{\delta^2}{2\sigma_R^2 (M)}\right)\;\;\;.
\end{equation}

To obtain the mass spectrum of PBHs one must introduce the $M_{BH}$ variable.
Besides, we will use the variable $\delta '$, connected with $\delta$
by the relation
\begin{equation}
\delta ' = \delta \left( \frac{M}{M_i} \right) ^{2/3}\;\;.
\end{equation}
Here, $M_i$ is the horizon mass at the moment of a beginning of the growth of density fluctuations.
This new variable is the density contrast at the moment of the collapse.
The new distribution function is 
\begin{equation}
\label{*}
n_{BH}(M_{BH},\delta ') = n(M_{BH},\delta ) \frac{d\delta}{d\delta '}
\frac{d M}{d M_{BH}}\;\;.
\end{equation}
Further, we assume that there is some functional connection between $M_{BH}$, $M$ and $\delta '$:
\begin{equation}
\label{0}
M_{BH}=f(M,\delta ')\;\;.
\end{equation}
In this case one can rewrite Eq.(\ref{*}) in the form
\begin{equation}
\label{29}
n_{BH} (M_{BH},\delta ') = n\left(M,\delta ' \left(\frac{M}{M_i}\right)^{-2/3}\right)\cdot \left(\frac{M}{M_i} \right)^{-2/3}
\cdot \frac{1}{{d f(M,\delta ')}/{d M}} 
\end{equation}
and the PBH mass spectrum is given by the integral
\begin{equation}
\label{210}
n_{BH}(M_{BH}) = \int n_{BH}(M_{BH},\delta ') d\delta ' \;. 
\end{equation}
Now, to connect the mass spectrum with the spectral index one can use the relations 
\begin{equation}
\label{**}
\sigma_R = \sigma_H (M) \left(\frac{M}{M_i} \right)^{-2/3}\;\; , \;\; \sigma_H (M) \sim M^{\frac{1-n}{6}}.
\end{equation}
In these relations $\sigma_H (M)$ is the horizon crossing amplitude, i.e., the standard 
deviation of the density contrast in the perturbed region at the moment when this fluctuation
crosses horizon. The derivation of  Eqs.(\ref{**}) is given, for completeness, in 
Sec.\ref{sec:normalization}.

Using Eqs.(\ref{**}) in the expression (\ref{25}) for $n(M,\delta)$ one obtains
\begin{equation}
n(M,\delta) = -\frac{n+3}{6}\cdot \sqrt{\frac{2}{\pi}} \frac{\rho_i}{M_i^2} \frac{1}{\sigma_H}
\left(\frac{M}{M_i} \right)^{-4/3} \left|\frac{\delta^2}{\sigma_R^2} -1\right| e^{-{\delta^2}/{2\sigma_R^2}}\;\;.
\end{equation}
Substituting this expression in the r.h.s. of the Eq.(\ref{29}) and using Eq(\ref{210}) one has 
\begin{equation}
\label{213}
n_{BH}(M_{BH})= 
\frac{n+3}{6}\cdot\sqrt{\frac{2}{\pi}}\rho_i \int \frac{1}{M^2 \sigma_H} \left|\frac{{\delta '} ^2}{\sigma_H^2}-1 \right|
e^{-\frac{{\delta '} ^2}{2\sigma_H^2}} \frac{1}{d f(M,\delta ')/ d M} d{\delta '} \;\;.
\end{equation}
Values of $M$ in r.h.s of Eq.(\ref{213}) are expressed through $M_{BH}$, $\delta '$ by the relation (\ref{0}).

Formula (\ref{213}) is the final expression for the PBH mass spectrum, the main result of this Section. It is valid for any relation 
between PBH mass $M_{BH}$, mass of the original overdense region $M$ and the density contrast $\delta '$.   

In the particular case of a near critical collapse 
\begin{equation}
\label{214}
f(M, \delta ') = M_i^{1/3}M^{2/3} k (\delta ' -\delta_c )^{\gamma_k}\equiv \xi M^{1/3}_i M^{2/3} \; ,
\end{equation}
therefore
\begin{equation}
M=M_{BH}^{3/2}M_i^{-1/2}\xi^{-3/2}\;\; ; \;\; d f/d M =\frac{2}{3} M_i^{1/2}M_{BH}^{-1/2}\xi^{3/2}\;\;.
\end{equation}

Eq.(\ref{214}) can be rewritten in the form, derived in the ref.~\cite{8}, using the connection between $M$ and the horizon mass
$M_{h}$ (which is equal to a fluctuation mass at the moment when the fluctuation crosses horizon):
\begin{equation}
\label{eee}
M_h=M_i^{1/3}M^{2/3}\;\;.
\end{equation}

The resulting formula for the spectrum,
\begin{equation}
\label{eq}
n_{BH}(M_{BH}) =
\frac{n+3}{4} \sqrt{\frac{2}{\pi}} \rho_i \sqrt{M_i} M_{BH}^{-5/2} \int^1_{\delta_c} \frac{1}{\sigma_H} \left| \frac{\delta '{}^2}
{\sigma_H^2}-1 \right| e^{-\frac{\delta^2}{2\sigma_H^2}}\xi^{3/2}d \delta ' \;\;,
\end{equation}
coincides, as can be easily proved, with the expression derived in ref.~\cite{9}.

The mass spectrum formula for the  Carr-Hawking collapse can be obtained analogously, with the relation 
$f(M,\delta ')= \gamma^{1/2}M_i^{1/3}M^{2/3}$ (see Eq.(\ref{c5}) below), or from Eq.(\ref{eq}) using the substitutions
\begin{equation}
\gamma_k\to 0\;\;,\;\;k\to \gamma^{1/2}\;\;,\;\;\delta_c\to\gamma
\end{equation}
and the approximate relation
\begin{equation}
\int^1_{\gamma} d\delta ' \left(\frac{\delta '{}^2}{\sigma_H^2} -1 \right) e^{-\frac{\delta '{}^2}{2\sigma_H^2}}
\approx \gamma e^{-\frac{\gamma^2}{2\sigma_H^2}}\;\;.
\end{equation}
In this case one obtains the expression derived in ref.~\cite{7}:
\begin{equation}
\label{220}
n_{BH}(M_{BH})=\frac{n+3}{4}\sqrt{\frac{2}{\pi}}\gamma^{7/4}\rho_iM_i^{1/2}
M_{BH}^{-5/2}\sigma_H^{-1}\exp\left(-\frac{\gamma^2}{2\sigma_H^2}\right)\;.
\end{equation}
One can see from Eqs.(\ref{eq}) and (\ref{220}) that the PBH mass spectrum following from
the Press-Shechter formalism has quasi power form in both considered 
cases $\left(\sim M_{BH}
^{-5/2}\right)$. In Carr-Hawking case $M_{BH}^{min}\sim M_h$ , in contrast 
with this in the critical collapse case it is possible that $M_{BH}\ll M_h$ .

\section{Formula for neutrino diffuse background from PBHs}
\label{sec:neutbackgr}
The starting formula for a calculation of the cosmological background 
from PBH evaporations is \cite{10}
\begin{equation}
\label{31}
S(E)=\int n_{com}\frac{1}{4\pi a_0^2\rho^2_{\mathstrut}}f\left(E(1+z)
\right)dV_{com}\;\;\;.
\end{equation}
Here , $n_{com}$ is the comoving number density of the sources (in our case
the source is an evaporating PBH of the definite mass $m$ ), $a_0$ is the 
scale factor at present time, $t=t_0$, $f(E)$ is a differential energy 
spectrum of the source radiation. $V_{com}$ is a comoving volume of the
space filled by sources, therefore 
\begin{equation}
\label{32}
dV_{com}=a_0^3\frac{\rho^2 d\rho}{\sqrt{1-k\rho^2}}d\Omega\;\;\;.
\end{equation} 
Here, $k$ is the curvature coefficient,
and $\rho$ is the radial comoving coordinate. Using the change of the variable,
\begin{equation}
\label{33}
\frac{d\rho}{\sqrt{1-k\rho^2}}=\frac{dt}{a}\;\;\;,
\end{equation}
the comoving number density can be expressed via the initial density $n_i$,
\begin{equation}
\label{34}
n_{com}=n_{phys}(t_0)=n_i\left(\frac{a_i}{a}\right)^3\left(\frac{a}{a_0}\right)
^3=n_i\left(\frac{a_i}{a_0}\right)^3\;\;\;\;.
\end{equation}
Substituting Eqs. (\ref{32}) - (\ref{34}) in Eq. (\ref{31}) one obtains
\begin{equation}
S(E)=n_i\int dt\frac{a_0}{a}\left(\frac{a_i}{a_0}\right)^3 f\left(
E(1+z)\right)\;\;\;.
\end{equation}
In our concrete case the source of the radiation is a Hawking evaporation:
\begin{equation}
\label{36}
n_i f\left(E(1+z)\right) = \int dm\, n_{BH}(m,t)f_H\left(E(1+z),m\right)\;\;\;.
\end{equation}
Here , $n_{BH}(m,t)$ is the PBH mass spectrum at any moment of time,
$f_H(E,m)$ is the Hawking function \cite{2},
\begin{equation}
\label{37}
f_H\left(E,m\right)=\frac{1}{2\pi\hbar}\frac{\Gamma_s(E,m)}{exp\left(\frac{8\pi
GEm}{\hbar c^3}\right)-(-1)^{2s}}\;\;\;.
\end{equation}
Here, $E$  is the energy of an evaporated particle (it lies in the interval [$0$,$\infty$] 
for massless particles), $\Gamma_s(E,m)$ is the coefficient of the absorption by a black hole
of a mass $m$, for an particle having spin s and energy $E$.

In our calculations (we considered only the simplest case of a nonrotating uncharged
hole) we used the approximation when $\Gamma_s$ depends only on $mE$ (as is the case
for massless particles), i.e. we ignore the decrease of $\Gamma_s$ at nonrelativistic 
energies of evaporated particles. But, working in this massless limit, we nevertheless
used in a low $mE$ region the exact forms of $\Gamma_s$, which are obtained numerically
(concretely, we used the plots for absorption cross section, $\sigma\approx\Gamma_s/{E^2}$,
as a function of $mE$, for $s=1/2$ and $s=1$, from the Fig.1 of work \cite{15}), rather than the low-energy
limits obtained analytically in works \cite{11,31}.   

The initial spectrum of PBHs is given by Eq. (\ref{220}). The minimum value of PBH mass 
in the specrtum can be obtained from the following considerations. It  is known \cite{5} 
that the condition of a recollapse of the perturbed region is the following: radius of a
maximum expansion of this region, $R_c$, must obey the inequality
\begin{equation}
\label{c1}
R_c > \gamma^{1/2} R_h = \gamma^{1/2} (ct_c) \;.
\end{equation}
Here, $R_h$ is the horizon radius at $t_c$, the moment of time, when the radius of 
the perturbation region is equal to maximum one. It is known also \cite{5}  that 
$t_c$ and $t_i$, the initial moment of time, are connected by the relation 
\begin{equation}
\label{c2}
\frac{t_i}{t_c}\sim \delta_i\;,
\end{equation}  
where $\delta_i$ is the initial density contrast for which there is the condition
following from Eq.(\ref{c1}):
\begin{equation}
\label{c3}
\delta_i\agt \gamma \left(\frac{M}{M_i}\right)^{-2/3}.
\end{equation}
Using Eq.(\ref{c1}) and the formula for horizon mass at $t_i$, 
\begin{equation}
\label{c4}
M_i=\frac{4}{3} \pi (c t_i)^3 \rho (t_i)\;\;,
\end{equation}
one obtains the connection between $M_{BH}$ and  $t_c$:
\begin{equation}
\label{c5}
M_{BH}=\frac{4}{3}\pi R_c^3 \rho (t_c) = \gamma^{3/2} M_i \frac{t_c}{t_i} \; .
\end{equation}
The use of Eqs.(\ref{c2},\ref{c3}) in Eq.(\ref{c5}) leads to the relation \cite{7}
\begin{equation}
\label{c5a}
M_{BH}=\gamma^{1/2} M_i^{1/3} M^{2/3} \; .
\end{equation}
From Eq.(\ref{c5a}) one follows that the minimum value of a PBH mass in the PBH mass spectrum
is given by the simple formula
\begin{equation}
\label{c6}
M_{BH}^{min}= \gamma^{1/2} M_i \; .
\end{equation} 
Strictly speaking, the minimum value of $\frac{t_c}{t_i}$ is equal to $1$ (corresponding
to $\delta_i^{max}=1$). Therefore, the absolute minimum value of PBH mass, following
from Eq.(\ref{c5}), is equal to $\gamma^{3/2} M_i$. However, the use of such a value     
as a border value of the PBH mass spectrum is clearly inconsistent. It contradicts with
the relation (\ref{c5a}) and, in general, with the fact that, due to the exponential damping \cite{7},
the bulk of the PBH intensity is determined just by the minimum value of $\delta_i$ (given by Eq.(\ref{c3})).

To take into account the existence of the minimum we must add to the initial spectrum 
expression the step factor $\Theta (M_{BH}-\gamma^{1/2}M_i)$. 
The connection of the initial 
mass value $M_{BH}$ and the value at any moment $t$ is determined by the solution 
of the equation \cite{11}
\begin{equation}
\label{38}
\frac{d m}{d t}=-\frac{\alpha (m)}{m^2}\;\;.
\end{equation}
The function $\alpha (m)$ accounts for the degrees of freedom of evaporated particles
and determines the lifetime of a black hole. In the approximation $\alpha = const$
the solution of Eq.(\ref{38}) is 
\begin{equation}
\label{310}
M_{BH}\cong\left(3\alpha t+m^3\right)^{1/3}\;\;\;.
\end{equation}
This decrease of PBH mass leads to
the corresponding evolution of a form of the PBH mass spectrum. At any moment one has
\begin{equation}
\label{311}
n_{BH}(m,t)dm=\frac{m^2}{\left(3\alpha t+m^3\right)^{2/3}}n_{BH}\left((3\alpha
t+m^3)^{1/3}\right)\times
\Theta\left[m-\left((\gamma^{1/2}M_{i})^3-3\alpha t\right)^{1/3}\right]dm.
\end{equation}
Substituting Eqs. (\ref{37}), (\ref{311}) in the integral in Eq.(\ref{36}) , we obtain the
final expression for the spectrum of the background radiation:
\begin{eqnarray}
\label{28}
S(E)=\frac{c}{4\pi}\int dt \frac{a_0}{a}\left(\frac{a_i}{a_0}\right)^3
\int dm \frac{m^2}{(3\alpha t+m^3)^{2/3}}n_{BH}\left[\left(3\alpha t+m^3\right)
^{1/3}\right]\times\nonumber\\
\\
\Theta\left[(m-\left((\gamma^{1/2}M_i)^3-3\alpha t\right)^{1/3}\right]
g_{\nu} f_H(E(1+z),m)\;\;\;.\nonumber
\end{eqnarray} 
Here, $g_{\nu}$ is a number of degrees of freedom. In the following we will interest 
in ($\nu_{e}+\tilde \nu_e$) - spectrum, so $g_{\nu}=2$.

One should note that the corresponding expressions for the spectrum in
refs.~\cite{12,13} contain the factor $\left(\frac{a_i}{a}\right)^3$ instead of the
correct factor $\left(\frac{a_i}{a_0}\right)^3$. It leads to a strong 
overestimation of large $z$ contributions in $S(E)$ (see below, Fig.\ref{fig:fig3}).

It is convenient to use in Eq.(\ref{28}) the variable $z$ instead of $t$. 
In the case of flat models with nonzero cosmological constant one has 
\begin{equation}
\label{saw}
\frac{dt}{dz}=-\frac{1}{H_0(1+z)}\left(\Omega_{m}(1+z)^3+\Omega_{r}(1+z)^4
+\Omega_{\Lambda}\right)^{-1/2}\;\;\;,
\end{equation}
\begin{equation}
\Omega_{r}=(2.4\cdot 10^4h^2)^{-1}\;\;\;\;,\;\;\;\Omega_{m}=1-
\Omega_{r}-\Omega_{\Lambda }\;\;\;.
\end{equation}
The factor $(\frac{a_i}{a_0})^3$ can be expressed through the value of $t_{eq}$, the moment 
of matter-radiation density equality:
\begin{equation}
\label{sdw}
\left(\frac{a_i}{a_0}\right)^3\simeq(1+z_{eq})^{-3}\left(\frac{t_i}{t_{eq}}
\right)^{3/2}=\left(\frac{2}{3}(2-\sqrt{2})\right)^{-3/2} H_0^{3/2}\left(2.4
\cdot10^4h^2\right)^{-3/4}t_i^{3/2}\;\;\;.
\end{equation}
The constant $H_0$, entering the Eqs.(\ref{saw},\ref{sdw}) is connected with $h$ by 
definition:
\begin{equation}
\label{sfe}
H_0=100 h \text{ km} \text{ s}^{-1} \text{ Mpc}^{-1}\;.
\end{equation}
Throughout the paper we suppose that $h=0.67$. One can see from Eqs.(\ref{sdw},\ref{sfe}) 
that the factor $\left(\frac{a_i}{a_0} \right)^3$
does not depend on $h$ and $\Omega_m$, as it must be.
  
Integrating over PBH's mass in Eq.(\ref{28}), one obtains finally, after the change of 
the variable $t$ on $z$, the integral
over~$z$:
\begin{equation}
\label{313}
S(E)=\int \,F(E,z) d\log_{10} (z+1)\;\;\;.
\end{equation}

In analogous calculations of the photon diffuse background integral over $z$ in the expression
for $S(E)$ is cut off at $z=z_0\approx 700$ because for larger $z$ the photon
optical depth will be larger than unity \cite{14}. In contrast with this , interactions
of neutrinos with the matter can be neglected up to very high values of $z$.
Therefore the neutrino diffuse background from PBH evaporations is much more
abundant.
 The neutrino absorption effects are estimated below, in Sec.\ref{sec:constr}.

The evaporation process of a black hole with 
not too small initial mass is almost an explosion. So, for a calculation of spectra of evaporated particles 
with acceptable accuracy it is enough to know the value of $\alpha$ for an initial
value of the PBH mass only. Taking this into account and having in mind the steepness of the PBH mass spectrum, 
we use the approximation
\begin{equation}
\alpha (m)=\alpha (M_{BH}^{min} )=\alpha(\gamma^{1/2} M_i)\;\;,
\end{equation}
and just this value of $\alpha $ is meant in the expressions (\ref{310})-(\ref{28}).

The very detailed calculation of the function $\alpha (m)$ was carried out
in the works \cite{15,16}. Here we use the simplified approach in which $\alpha (m)$
is represented by the dependence
\begin{equation}
\label{36a}
\alpha=\alpha_0 + \sum\limits_i a_i\cdot 10^{25}\cdot \Theta(b_i-log_{10}(m)).
\end{equation}
\begin{table}[htb] 
\caption{Coefficients $a_i$ and $b_i$ in Eq.(\ref{36a})  \label{tab:t1}}
\begin{tabular}{@{\hspace{.4in}}lllllllllll@{\hspace{.4in}}}

$i$   & $\mu$     & u,d       &g      & s      & $\tau$      & c      &  b      & $W^{\pm}$       & $Z^0$ & t       \\
\hline
$a_i$ & 3.12      &  9.36     &4.96   & 9.36   &  3.12       & 9.36   &  9.36   &  1.86           & 0.93    & 9.36  \\
$b_i$ & 14.59     &  14.13    &13.97  & 13.89  &  13.35      & 13.32  &  12.91  &   11.87         & 11.85   & 11.39 \\ 

\end{tabular}
\end{table} 

Here, $\Theta (x)$ is the step function. Coefficient $\alpha_0$ gives the summary
contribution of $e^{+}$, $e^{-}$, $\nu$ and $\gamma$ and is equal to $8.42\cdot 10^{25}\; g^3 sec^{-1}$
\cite{11,16}. All other coefficients are collected in the Table~\ref{tab:t1}.

The coefficients $b_i$ determine
the value of the PBH mass beginning from which particles of $i$-type can be evaporated. This value, $M_{BH}^{b_i}$,
is obtained from the relations 
\begin{equation}
\label{arr}
\left(\frac{10^{13}g}{M_{BH}^{b_i}}\right)\text{ GeV}\cong k T_{BH}^{b_i} \cong \frac{m_i}{\kappa (s_i)}\;,
\end{equation}
\begin{equation}
\label{arr1}
\kappa (\frac{1}{2}) \approx 4\;\;\;,\kappa (1) \approx 6\;. 
\end{equation}
So, 
\begin{equation}
b_i =log_{10} M_{BH}^{b_i}\;\;.
\end{equation}
Writing Eqs.(\ref{arr},\ref{arr1}) we took into account that the evaporation of massive particles 
becomes essential once the peak value in the energy distribution exceeds the particle rest 
mass. The peak value depends on the particle spin (see, e.g., \cite{15}):
\begin{equation}
\label{arr2}
E_{peak}(s=1/2)\approx 4 k T_{BH}\;\;\;;\;\;\;E_{peak}(s=1)\approx 6 k T_{BH}\;.
\end{equation}

\begin{figure}[!t]
\epsfig{file=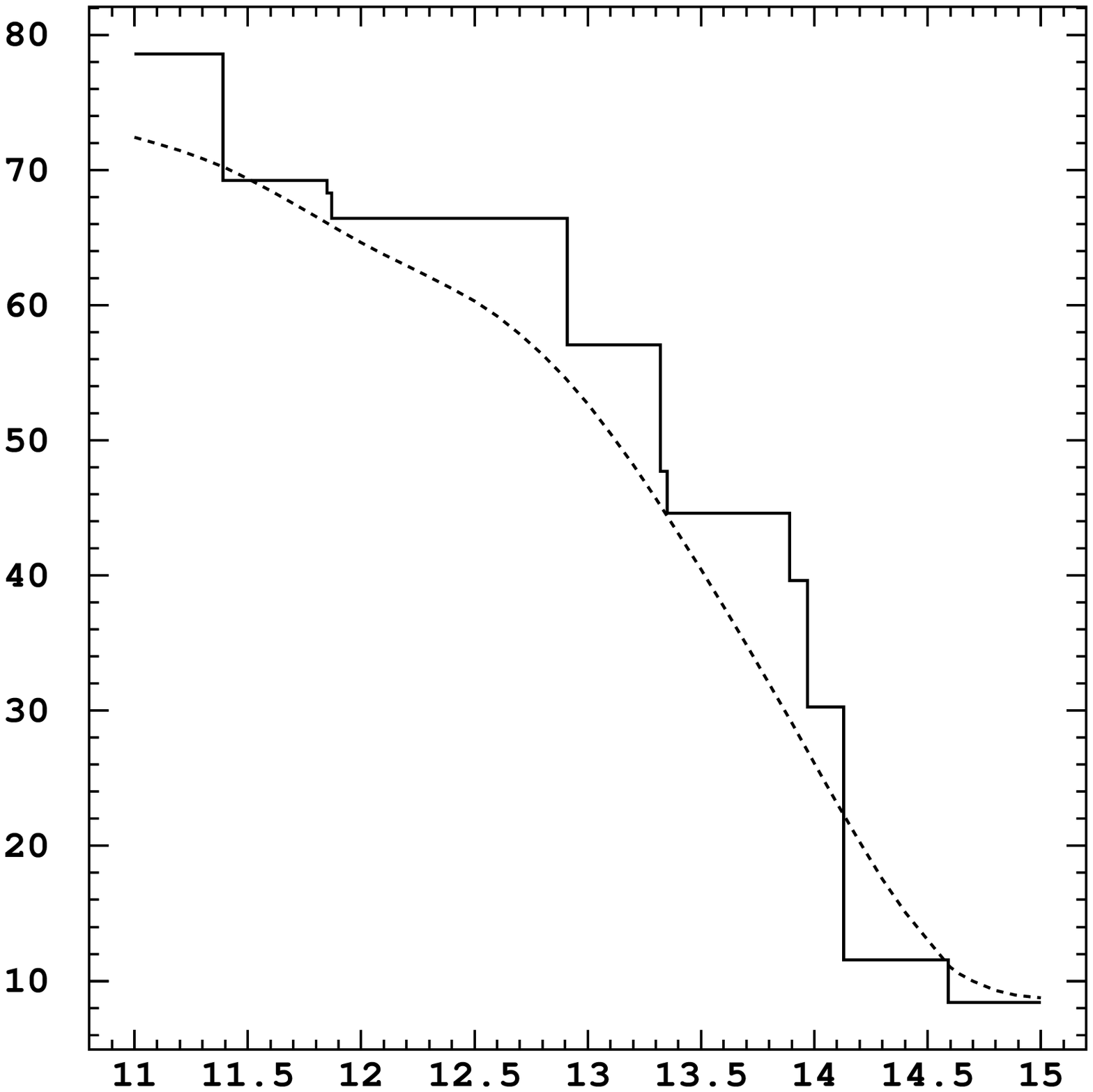,width=\columnwidth}
\caption{The function $\alpha (m)$ counting the degrees of freedom of the PBH radiation ($m$ is
an instantaneous value of the PBH mass).
Solid line is the present result, dashed line is drawn using Eq.(7) of Ref.[17].}
\label{fig:fig1}
\end{figure}

The values of masses of (u,d,s,c,b,g,$\tau$)-particles used in our calculations of $b_i$
are the same as in work \cite{16}, top quark mass was taken equal to $170\text{GeV}$. 
The resulting function $\alpha (m)$ is shown on Fig.\ref{fig:fig1}. For comparison, the corresponding dependence
from the work \cite{16} (drawn using the Eq.7 of \cite{16}) is also shown. 

According to Eq.(\ref{c6}) the minimum value of PBH mass in the mass spectrum is about the horizon mass
$M_i$. In turn, the value of $M_i$ is connected with $t_i$, the moment of time just after reheating, from
which the "normal" radiation era and the process of PBHs formation start, 
\begin{equation}
\label{g622}
M_i=\frac{1}{8}\frac{M_{pl}}{t_{pl}} t_i\;.
\end{equation}  
The corresponding reheating temperature is given, approximately, by the relation (see Eq.(\ref{l1})) 
\begin{equation}
k T_{RH}\simeq \left(\frac{t_i}{1s}\right)^{-1/2}\text{MeV}\;.
\end{equation}
If $k T_{RH}$ values lie in the interval $(10^8 - 10^{10})\text{ GeV}$ (just for this interval we obtain the spectral index
constraints in the present paper), the corresponding $M_i$ values lie in the mass interval $(10^{11}-10^{15})\text{ g}$.
So, as one can see, we really need for our aims the all information about $\alpha$-parameter, that contains in Fig.\ref{fig:fig1},
including the highest values of $\alpha$ at $m\sim 10^{11}\text{ g}$.

The formula (\ref{28}), as it stands, takes into account the contribution to the neutrino background solely from
a direct process of the neutrino evaporation. If we suppose that particles evaporated by a black hole 
propagate freely, the calculation of other contributions to the neutrino background can be performed using our
knowledge of particle physics \cite{15,16}.

On Fig.\ref{fig:fig2} the typical result of our calculation of instantaneous neutrino spectra from evaporating black hole is shown.
The spectrum of the straightforward (direct) $(\nu_e+\tilde\nu_e)$ - emission is given by the Hawking function $f_H (E,m)$, Eq.(\ref{37}).
The $(\nu_e + \tilde\nu_e)$ spectrum arising from decays of $(\mu^{+} + \mu^{-})$ evaporated directly is calculated by the formula
\begin{equation}
f^{(\mu)}(E,m) = \int \Theta (b_{\mu} -\log_{10} m) g_{\mu}  f_H (E_{\mu},m)\frac{d n_{\nu} (E_{\mu},E)}{d E}d E_{\mu}\;\;,
\end{equation} 
where $d n^{\mu}_{\nu}/d E$ is the neutrino spectrum in a $\mu$ - decay, $g_{\mu}=4$. To evaluate the electron neutrino spectrum resulted from fragmentations of 
evaporated quarks, we used the simplest chain
\begin{equation}
(u,d)-\mbox{quarks}\;\; \longrightarrow\;\; \pi\;\; \longrightarrow\;\; \mu\;\; 
\longrightarrow\;\; \nu_e\;\;,
\end{equation}
and the formula 
\begin{equation}
f^{(q)} (E,m)= \int \Theta (b_q - \log_{10} m) g_{q} f_H (E_q,m)\frac{d n^q_{\pi}(\xi)}{d \xi}\cdot\frac{d \xi}{d E_q}\cdot \frac{d n_{\mu}^{\pi}(E_{\mu},E_{\pi})}{d E_{\mu}}
\cdot\frac{d n^{\mu}_{\nu}(E_{\mu},E)}{d E} d E_{q} d E_{\pi} d E_{\mu} \;\;.
\end{equation}
Here, $g_{q}=24$ for a summary contribution from $u,\; \tilde u,\; d,\; \tilde d$ - quarks,  
${d n^q_{\pi}}/{d \xi}$ is the $q\to \pi$ fragmentation function, for which the simple parametrization was taken:
\begin{equation}
\label{sqw}
\frac{d n^q_{\pi}}{d \xi} =\frac{15}{16}(\xi - 1)^2 \xi^{-3/2}\;\;,\;\; \xi=\frac{E_{\pi}}{E_q}\;\;,
\end{equation} 
and ${d n_{\mu}^{\pi}}/{d E_{\mu}}$ is the $\mu$ - spectrum in
a decay $\pi \to \mu + \nu_{\mu} $.

Analogous formulas are used for calculations of the neutrino spectrum from other channels of the neutrino production, 
for instance from decays of evaporated $W$-bosons
($W\to e + \nu_e\;\;,$ $W\to \mu \to \nu_e$).

The formula (\ref{sqw}) for the fragmentation function was suggested in the paper
\cite{32}. It leads to a simple multiplicity growth law of $\sqrt{E}$ (such a law follows
from a naive statistical model of an jet fragmentation ). It had been shown in  
ref.~\cite{32}  that Eq.(\ref{sqw}) fits rather well the PETRA data and can be used instead of
the more ingenious formula based on the leading logarithm approximation
of QCD. In turn, it was shown recently \cite{33} using the programme HERWIG \cite{34} that
the parametrizations of ref. \cite{32}, being rather good at low $\xi$, lead to an significant
overestimation of the yield of large $\xi$ final states. This fact as well as the
simplicity of Eq.(\ref{sqw}) were decisive for our choice of the fragmentation function
because, as we will show, we need just the upper estimate of the yield of quark 
fragmentations.

In all calculations in the present paper we neglected the contribution to neutrino fluxes
from neutron decays in the jets. These decays give the distinct bumps in instantaneous $\nu \tilde \nu$-%
spectra at $E\sim 1\mbox{ MeV}$ \cite{15}. In general, neutrinos of such low energies have too small cross sections
for neutrino-nucleus and neutrino-nucleon interactions (see sec.\ref{sec:constr}) and are 
inessential for our main aim: the constraining of the spectral index (neutrinos with energies
$\sim 10-100\mbox{ MeV}$ are mainly responsible for this). Besides, we will see at Sec.\ref{sec:spectra}
that at large values of one of the parameters of our approach (reheating temperature $T_{RH}$)
these neutrinos are not noticeable in the summary neutrino background
even at $E\sim 1\mbox{ MeV}$ because of strong redshift effects.

In order to  check the  accuracy of  our  calculations of  instantaneous neutrino spectra 
we calculated for  one  value of a black  hole temperature the summary  neutrino spectrum
(excluding secondary $\nu_{\tau}$ only). The result is shown on Fig.\ref{fig:fig11} together
with the corresponding curve from ref.\cite{15}. One can see from this figure that in the
practically interesting region of neutrino energies ($E  >  100 \text{ MeV}$)  our 
result differs from the Monte-Carlo result of ref.\cite{15}  not more  than on a factor of 2.  

The relative contribution of different channels to the total $\nu_e$ spectrum strongly depends on 
the black hole temperature. One can see from Fig.2, that at high temperature decays of massive particles evaporated by the black hole
become rather important at high energy tail of the spectrum (if the corresponding branching ratios are not too small). 

The total instantaneous neutrino spectrum from a black hole evaporation is given by the sum
\begin{equation}
f(E,m)= g_{\nu} f_H(E,m)+f^{(\mu)}(E,m)+f^{(q)}(E,m)+ ...\;\;\;,
\end{equation}
and the total background neutrino spectrum is given by the same Eq.(\ref{28}), except the change $g_{\nu}f_H(E,m)\to f(E,m)$.

\section{Normalization of the perturbation amplitude}
\label{sec:normalization}

For numerical calculations of the neutrino background we need normalization
of the dispersion $\sigma_R$ entering the PBH mass spectrum formula. For this
aim we must consider the time evolution of this dispersion. The connection
of $\sigma_R (t)$ with a power spectrum $P(k,t)$ of the density perturbations is given by 
the expression
\begin{eqnarray}
\label{d1}
\sigma_R^2 (t) = \int\limits_0^\infty \frac{k^3}{2\pi^2} P(k,t) W^2 (kR)
\frac{d k}{k} \equiv\nonumber\\
\\
\int\limits_0^\infty \frac{k^4}{(aH)^4}\delta_H^2 (k,t) W^2 (kR)\frac{dk}{k} \; 
. \nonumber
\end{eqnarray}
Here, $W(kR)$ is the smoothing window function. 
The second line of this equation defines the (nonsmoothed) horizon crossing amplitude $\delta_H (k,t)$.
In general, it depends on the time and we need the initial amplitude $\delta_H (k,t_i)$,
whereas the normalization on COBE data (which had been performed, in particular, in the work \cite{44}) 
determines this amplitude at present time, $t=t_0$. Only in the particular case of a 
flat cosmological model with $\Lambda=0$ the horizon crossing amplitude is time-independent
and the normalization on COBE data gives $\delta_H (k,t_i)$ straightforwardly.

Assuming a power law of primordial density perturbations,
\begin{equation}
\label{d2}
P(k,t)\sim k^n,
\end{equation}
and choosing for $W(kR)$ the top-hat form,
\begin{equation}
\label{d3}
W(kR)=\frac{3 j_1 (kR)}{kR} ,
\end{equation}
one obtains from Eq.(\ref{d1})
\begin{equation}
\label{d4}
\sigma_R^2 (t)= C_n^2 \frac{k_{fl}^4}{(aH)^4}\delta_H^2 (k_{fl},t)
\equiv \frac{k_{fl}^4}{(aH)^4}\sigma_H^2 (k_{fl},t) \; .
\end{equation}
In Eq.(\ref{d4}), $k_{fl}$ is the comoving wave number, characterizing the
perturbed region,
\begin{equation}
\label{d5}
k_{fl}=\frac{1}{R}
\end{equation}
and $C_n$ is a constant which depends on the spectral index $n$. For
$n\simeq 1$ one has $C_n\simeq 4.8$ .
The fluctuation power per logarithmic interval, or variance, at present
time is given by the expression \cite{35}
\begin{equation}
\label{d6}
\frac{k^3}{2\pi^2}P(k,t_0)= \frac{1}{H_0^4}k^4 A_S^2 (k) \frac{g^2(\Omega_m)}
{\Omega_m^2}
\;.
\end{equation}

Here, $A_S(k)$ is an amplitude of the scalar perturbations, k-dependence of
which arises from a deviation from scale invariance ($A_S^2\sim k^{n-1}$).
The factor $g^2 (\Omega_m )$ takes into account the suppression of the growth
of density perturbations due to nonzero $\Omega_{\Lambda}$ (density 
perturbations stop growing once the universe becomes cosmological constant
dominated). The factor $\Omega_m^2$ in the denominator of Eq.(\ref{d6})
arises because the potential fluctuations are reduced by $\Omega_m$:
due to this, for fixed COBE normalization, the power spectrum of 
density perturbations must be, aside from other factors, enhanced
by $\Omega_m^{-2}$ \cite{36}.

For arbitrary moment of time in matter era the expression for the 
power spectrum is evident generalization of Eq.(\ref{d6}):      
\begin{equation}
\label{d14}
\frac{k^3}{2\pi^2}P(k,t)=\frac{k^4}{H_0^4}A_S^2(k)\left(\frac{g(a,\Omega_m)}
{\Omega_m}\right)^2\;.
\end{equation}

Here, the function $g(a,\Omega_m)$ accounting for the growth of density perturbations 
is given by the expression \cite{37}
\begin{equation}
\label{d10}
g(a , \Omega_m)= \frac{5\Omega_m}{2a}\frac{d a}{d\tau}\int\limits_0^a
\left(\frac{d a'}{d\tau}\right)^{-3} d a'  ,
\end{equation}
where $\tau\equiv H_0 t$. In our case 
\begin{equation}
\label{d11}
\frac{da}{d\tau}= \Omega_m\cdot \frac{1}{a} + \Omega_{\Lambda} a^2
\end{equation}
and
\begin{equation}
\label{d12}
g(a,\Omega_m)=\frac{5}{2}\Omega_m a^{-3/2}(\Omega_m+\Omega_{\Lambda}a^3)^{1/2}
\int\limits^a_0\frac{x^{3/2}dx}{(\Omega_m+\Omega_{\Lambda}x^3)^{3/2}} 
\; .
\end{equation}
The normalization of this function is such that 
\begin{equation}
\label{d12a}
g(a,1)=a\;.
\end{equation}
For the present time ($a=a_0=1$) one has 
\begin{equation}
\label{d13}
g(1,\Omega_m)=g(\Omega_m) \;,
\end{equation}
where $g(\Omega_m)$ is the function entering Eq.(\ref{d6}), as it must be.
The horizon crossing amplitude $\delta_H(k,t)$ defined
by the Eq.(\ref{d1}) is determined by the relation
\begin{equation}
\label{d15}
\delta_H^2(k,t)=\frac{(aH)^4}{H_0^4}A_S^2(k)\left(\frac{g(a,\Omega_m)}
{\Omega_m}\right)^2\;\stackrel{t\to t_0}{\longrightarrow}\; A_S^2(k)
\left(\frac{g(\Omega_m)}{\Omega_m}\right)^2
 \; .
\end{equation}
One can see that the time dependence of $\delta_H (k,t)$ arises just due to nonzero
value of $\Lambda$. If $\Lambda = 0$, one has from Eq.(\ref{d15}), using 
Eq.(\ref{d12a}),
\begin{equation}
\left.\delta_H^2 (k,t)\right|_{\Lambda = 0} = A_S^2 (k)\;.
\end{equation} 

Our normalization of perturbation amplitude on COBE data is based on two
inputs.

1. Connection between CMB fluctuations and scalar density fluctuations is 
described \cite{35} by the following relations (derived for a lowest-order
reconstruction of the inflationary potential):   
\begin{equation}
\label{d16}
S\equiv \frac{5C^2_S}{4\pi}= 0.104 f_S^{(0)}(\Omega_{\Lambda})A_S^2(k_*) \; ,
\end{equation}
\begin{equation}
\label{d16b}
f_S^{(0)} (\Omega_{\Lambda})= 1.04-0.82\Omega_{\Lambda}+2\Omega_{\Lambda}^2
 \; ,
\end{equation}
\begin{equation}
0.0\le \Omega_{\Lambda} \le 0.8 \; .
\end{equation}
Here, $C^S_2$ is the scalar contribution to the angular power spectrum of CMB
temperature fluctuations for $l=2$, $k_* \sim a_0 H_0 $ is the comoving wave 
number at the present horizon scale. The dependence $f_S^{(0)}$ on $\Omega_{\Lambda}$
is due to the integrated Sachs-Wolfe effect, i.e., due to the evolution of the
potentials from last scattering surface till the present time.

2. From 4-year COBE data \cite{38} one has (assuming that the scalar contribution
dominates over the tensor on the large scales) the following results for
small $l$ values \cite{39}:
\begin{equation}
\label{d17}
\left(\frac{l(l+1)C_l^S}{2\pi}\right)^{1/2}\simeq (1.03\pm 0.07)\cdot 10^{-5},
\end{equation}
\begin{equation}
\label{g001}
n=1.02\pm 0.24\;.
\end{equation}

From Eqs.(\ref{d15}-\ref{d17}) one obtains, finally, the normalization of the horizon crossing amplitude:
\begin{equation}
\label{d18}
\delta_H(k_*,t_0)\equiv \delta_H^{(COBE)}\simeq 2\cdot 10^{-5}\frac{1}
{\sqrt{f^{(0)}_{S}(\Omega_{\Lambda})}}\left(\frac{g(\Omega_m)}{\Omega_m}\right),
\end{equation}
\begin{equation}
A_S(k_*)=2\cdot 10^{-5}\frac{1}{\sqrt{f^{(0)}_S(\Omega_{\Lambda})}}.
\end{equation}
All the dependence of $\delta_H(k,t)$ on $k$ is contained in an amplitude
$A_S(k)$ and, according to our assumption, Eq.(\ref{d2}), is very simple:
\begin{equation}
\label{d19}
A_S(k)\sim k^{\frac{n-1}{2}}\;.
\end{equation}
Expressing the variable $k$ through $M_h$, horizon mass at the moment
when the scale $\lambda\sim\frac{1}{k}$ crosses the Hubble radius, one obtains
\begin{equation}
\label{d20}
A_S(M_h)\sim\left\{
          \begin{aligned}[l]
                M_h^{\frac{1-n}{4}}&\;,\;&\text{radiation era ,} \\
		& &                                            \\
                M_h^{\frac{1-n}{6}}&\;,\;&\text{matter era .}
          \end{aligned} 
	  \right.
\end{equation}
Writing Eqs.(\ref{d20}) we took into account that the comoving mass density 
in radiation era decreases with time \cite{46}, while in matter era it is
conserved.   
Now, using the Eq.(\ref{d15}) one can calculate the amplitude $\delta_H(k)$
at the moment of matter - radiation equality:
\begin{equation}
\label{d21}
\delta_H(M_{eq},t_{eq})= \frac{(a_{eq}H_{eq})^2}{H_0^2}A_S(M_{eq})\frac{g(a_{eq},
\Omega_m)}{\Omega_m}\;.
\end{equation}
Here, $M_{eq}$ is the horizon mass at $t_{eq}$.
Since $a_{eq}\ll 1$, one has 
\begin{equation}
\label{d22}
H_{eq}^2\simeq H_0^2\frac{\Omega_m}{a_{eq}^3}\;\;,\;\; 
g(a_{eq},\Omega_m)\simeq a_{eq}\;.
\end{equation}
Therefore, according to Eq.(\ref{d18}),
\begin{equation}
\label{d23}
\delta_H(M_{eq},t_{eq})= A_S (M_h)=2.0\cdot 10^{-5}\frac{1}
{\sqrt{f^{(0)}_S(\Omega_{\Lambda})}}\left(\frac{M_{eq}}{M_{h 0}}\right)^{\frac{1-n}{6}}\;.
\end{equation}
Here, $M_{h 0}$ is the present horizon mass. Using Eq.(\ref{d20}) one 
obtains the time-independent amplitude $\delta_H(M_h)$ in radiation era:
\begin{equation}
\label{d24}
\delta_H(M_h)= 2\cdot 10^{-5} \frac{1}{\sqrt{f_S^{(0)}(\Omega_{\Lambda})}}\left(
\frac{M_{eq}}{M_{h0}}\right)^{\frac{1-n}{6}}\left(\frac{M_h}{M_{eq}}\right)
^{\frac{1-n}{4}}\;.
\end{equation} 
The values of horizon masses $M_{eq}$ and $M_{h0}$ in Eq.(\ref{d24}) depend on $\Omega_m$:
\begin{equation}
\label{d25}
M_{eq}=\Omega_m^{-2} h^{-4}\cdot 1.5\cdot 10^{49}g\;\; ;\;\; M_{h0}=
\Omega_m h^{-1}\cdot 6\cdot 10^{55}\; g.
\end{equation}
The expression for $\sigma_H(M_h)$ differs from Eq.(\ref{d24}) only by
the numerical coefficient (see Eq.(\ref{d4})):
\begin{equation}
\label{d26}
\sigma_H (M_h)=C_n\delta_H (M_h) ,
\end{equation}
and $M_h$ now is a function of $k_{fl}$ .

One can see from Eqs.(\ref{d24},\ref{d25}) that the horizon crossing amplitude
$\sigma_H(M_h)$ , normalized on COBE data, depends on $\Omega_m$ only through the 
$\Omega_{\Lambda}$-dependence of the $f_S^{(0)}$ function:
\begin{equation}
\label{d27}
\sigma_H (M_h)\sim \frac{1}{\sqrt{f_S^{(0)}(\Omega_{\Lambda})}} M_h^{\frac{1-n}{4}}\;.
\end{equation}
The relations (\ref{d24}) and (\ref{d27}) are main results of this Section. 
The final expression for $\sigma_R(t)$ is, according to Eq.(\ref{d4}),
\begin{equation}
\label{d28}
\sigma_R(t)= \left(\frac{M_{hor}(t)}{M_{fl}(t)}\right)^{2/3}\sigma_H(M_h)\;.
\end{equation}
Here, $M_{fl}(t)$ is the fluctuation mass,
\begin{equation}
\label{d29}
M_{fl}(t)=\frac{4}{3}\pi\left(\frac{a(t)}{k_{fl}}\right)^3\rho(t)
\end{equation}
and $M_{hor}(t)$ is the horizon mass at a moment t. At the initial moment
of time one has ($M_{hor}(t_i)=M_i\;,\;M_{fl}(t_i)=M$)
\begin{equation}
\label{d30}
\sigma_R(t_i)\equiv \sigma_R (M)=\left(\frac{M_i}{M}\right)^{2/3}
\sigma_H(M_h)\;.
\end{equation}
The  expressions (\ref{**}) of Sec.\ref{sec:PBHsp} can be obtained from 
Eqs.(\ref{d27} , \ref{d30}) using the connection between $M$ , $M_i$ and 
$M_h$ for radiation era (Eq.(\ref{eee})).

\section{Neutrino background spectra}
\label{sec:spectra}

A precise calculation of the PBH neutrino background must include also a taking into account the neutrino absorption during a travelling in the space.
In the case of the photon background the most important absorption process is a pair production on neutral matter \cite{14} due to which the photons from 
PBHs evaporated earlier than $z\approx 700$ are absent today. In our case, the analog of an optical depth of the universe for the neutrino emitted at
a redshift z and having today an energy $E$ is given by the integral\footnote{ In formulas of Secs.\ref{sec:spectra},
\ref{sec:constr} we use the convention $\hbar=c=k=1$.}
\begin{equation}
\label{RRR}
\tau (z,E) =  \int\limits_0^z\sigma\left( E(1+z')\right)\cdot n(z') \frac{d t}{d z'} d z'\;\;.
\end{equation}
Here, $\sigma (E)$ is the neutrino interaction cross section, $n(z)$ is a number density of the target particles. 

Two processes are potentially
"dangerous": neutrino-nucleon inelastic scattering growing linearly with an energy, and annihilations with neutrinos of the relic background
\begin{equation}
\nu_e + N \to e^{-} + anything,
\end{equation}
\begin{equation}
\nu_e + \tilde\nu_e (relic) \to \sum\limits_i (f_i +\tilde f_i).
\end{equation}
Here, $f_i$ are charged fermions (leptons and quarks). As is known, the relic neutrino background exists, with a Planck 
distribution, beginning from the epoch of neutrino decoupling ($z\approx 10^{10}$).

The cross section of the neutrino-nucleon inelastic scattering can be
approximated (in the neutrino energy interval from $E\sim 1 \mbox{GeV}$ to $E\sim
10^5 \mbox{GeV}$) by the simple formula:
\begin{equation}
\label{f1}
\sigma_{\nu N}(E)\approx 0.5\cdot 10^{-38}\left(\frac{E}{\mbox{GeV}}\right)
 cm^2 \; .
\end{equation}
At $E\agt 10^5\text{ GeV}$ the cross section grows with energy more slowly, so
the  use of Eq.(\ref{f1}) in this region leads to the overestimation of the cross section value.
But we will use this formula for the entire interval of neutrino energy, having
in mind that practically we need only the upper estimate of this cross  
section (because, as we will see, the contribution of this channel to the
total absorption factor is small).

The number density of the target nucleons can be estimated using the existing
constraints on fraction of critical density in baryons \cite{47}:
\begin{equation}
0.004 < \Omega_B h^2 < 0.021. 
\end{equation}
From these inequalities and for $h=0.67$ one obtains for the present nucleon 
number density :
\begin{equation}
n_N^0=\frac{n_N(z)}{(1+z)^3}=(4\cdot 10^{-8} \div 2\cdot 10^{-7})cm^{-3}.
\end{equation}
In numerical calculation we used the value $10^{-7}$ for $n_N^0$. The
$\tau_{\nu N}$-function is given by the simple integral:
\begin{equation}
\tau_{\nu N}(E,z)\approx 0.5\cdot 10^{-38}\left(\frac{E}{\mbox{GeV}}\right)
\cdot 10^{-7}\cdot\int\limits_0^z (1+z)^4 \frac{d t}{d z'} {d z'}
 \; .
\end{equation}

The expression (\ref{RRR}), as it stands, is valid only in a case when a target 
particle is at rest. In the case of an absorption through annihilations
of evaporated neutrinos with relic neutrinos the current expression for
the $\tau_{\nu \tilde {\nu}}$ - function is (see, e.g., ref.~\cite{30})
\begin{equation}
\label{w58}
\tau_{\nu \tilde {\nu}}(E,z)=\int\limits_0^z \langle (1-\cos \theta_{\nu 
\tilde {\nu}}) \sigma_{\nu \tilde {\nu}} \left(E(1+z'),E_{\nu}(1+z'),
\theta_{\nu \tilde {\nu}} \right)\rangle n_{\tilde {\nu}} (z')\cdot 
\frac{d t}{d z'} d z'
 \; .
\end{equation}

The angular brackets indicate an average over the relic antineutrino
energy distribution and over the angle $\theta_{\nu \tilde {\nu}}$
between the two colliding particles in the cosmic frame.

The cross section of the $\nu \tilde {\nu}$ - annihilation is expressed by the formula \cite{30,41}
\begin{equation}
\label{sxc} 
\sigma_{\nu \tilde {\nu}}(E,E_{\tilde {\nu}},\theta_{\nu \tilde {\nu}})=
\frac{G_F^2}{4\pi}\left[ N_{eff}^{NC}(s) + N^{CC}_{eff}(s)\right]s,
\end{equation}
\begin{equation}
\label{sdc}
s=2 E E_{\tilde {\nu}}(1-\cos \theta_{\nu \tilde {\nu}}).
\end{equation}

Here, $s$ is a square of the summary energy of colliding particles in the
center of mass system, the coefficients $N_{eff}$ are the effective numbers
of annihilation channels (for neutral and charged currents).

Formula (\ref{sxc}) is valid in the energy region before the $Z$ boson pole,
i.e., when
\begin{equation}
s \ll M_Z^2.
\end{equation}

As one can see from Eq.(\ref{sdc}), maximum values of $s$, for which we must
know the cross section, are about $E T^0_{\tilde \nu} (1+z)^2$ (where $T^0_
{\tilde \nu}$ is the present temperature of the relic antineutrino gas,
$z$ is the upper limit of the integral in Eq.(\ref{w58}), and $E$ is the 
neutrino energy today). So, the cross section formula (\ref{sxc}) can be used if
\begin{equation}
\label{ddf}
z\ll\left(\frac{M_Z^2}{E\cdot T_{\tilde \nu}}\right)^{1/2}\approx
3\cdot 10^8 \left(\frac{E}{\mbox{GeV}}\right)^{-1/2}.
\end{equation}
We will see from the numerical results that this condition is satisfied in
all the practically important cases.

The coefficients $N_{eff}$ are given by the following expressions:
\begin{equation}
N_{eff}^{NC} (s)=\sum_f \Theta (\frac{s}{4} - 
m_f^2)\cdot\frac{2}{3}n_f \left(1-8 t_{3f}q_f\sin^2\theta_W + 8 q_f^2
\sin^4\theta_W\right),
\end{equation}
\begin{equation}
N_{eff}^{CC}= \Theta (\frac{s}{4} - m_e^2)\cdot\frac{16}{3}\sin^2\theta_W.
\end{equation}
Here, $n_f$ is the number of colours (1 for leptons, 3 for quarks ), $t_{3f}$
and $q_f$ are the third component of the weak isospin and the electric charge
( in units of the positron charge ), respectively, $\theta_W$ is the 
electroweak mixing angle ($\sin^2 \theta_W\approx 0.23$), $\Theta (x)$ is 
the step function.

Using the relations
\begin{equation}
\langle (1-\cos\theta_{\nu \tilde \nu})^2 E_{\tilde \nu}\rangle=
\frac{4}{3} \rho_{\tilde \nu}\frac{1}{n_{\tilde \nu}},
\end{equation}
\begin{equation}
\langle \frac{s}{4} -m_f^2 \rangle \cong E T_{\tilde \nu}^{0}-m_f^2\;,
\end{equation}
where $\rho_{\tilde \nu}$ is the relic antineutrino energy density, one 
obtains the final expression for $\tau_{\nu \tilde \nu}$:
\begin{equation}
\tau_{\nu \tilde \nu}(E,z)\cong\frac{G^2}{4\pi}\cdot\frac{8}{3}\rho_{\tilde \nu}^0
E\int\limits_0^z (1+z')^5 
\frac{d t}{d z'} 
N_{eff} \left(E T_{ 
\tilde \nu} ^0 (1+z')^2\right) d z' ,
\end{equation}
\begin{equation}
N_{eff}=N_{eff}^{NC}+N_{eff}^{CC}.
\end{equation}
Here, $\rho_{\tilde \nu}^{0}$ is the present electron antineutrino energy density,
\begin{equation}
\rho_{\tilde \nu}^0= \frac{7}{8}\cdot \frac{\pi^2}{30}(T^0_{\tilde \nu})^4\;.
\end{equation} 

Results of the calculations (with $h=0.67$, $\Omega_{\Lambda}=0$) of 
$\tau_{\nu N} $ and $\tau_{\nu \tilde \nu}$ are shown on Fig.\ref{fig:fig8}
for several values of neutrino energy. One can see, that the contribution
of $\nu N$ channel to the total $\tau$ is negligibly small everywhere. For
typical neutrino energy $\sim 100\text{ MeV}$ the absorption is essential
beginning from $z\sim 3\cdot 10^6$. A growth of the absorption factor with $z$ is 
very fast, therefore the redshifts for which the condition (\ref{ddf}) 
is violated are never really important in a calculation of the absorption.

Our calculation of neutrino spectra from evaporating PBHs contains two 
parameters: a spectral index $n$ and a time of an end of the inflation $t_i$
(which, by assumption, is a time when density fluctuations develop). We 
assume that at $t_i$ the universe has as a result of the reheating the 
equilibrium temperature $T_{RH}$. The connection of $T_{RH}$ and $t_i$
is given by the standard model:
\begin{equation}
\label{l1}
t_i=0.301 g_*^{-1/2}\frac{M_{pl}}{T_{RH}^2}\approx\frac{0.24}{(T_{RH}/1MeV)^2}
\;\;\;s
\end{equation}
($g_*\sim100$ is the number of the degrees of freedom in the early universe).

On Fig.\ref{fig:fig3} the integrand of the background spectrum integral, considered in 
Sec.\ref{sec:neutbackgr} 
is shown as a function of redshift $z$ for several values 
of the parameter $T_{RH}$. The inclusion of the absorption effects leads to an appearing 
of the additional factor in this integral:
\begin{equation}
\label{sdfr}
S(E)=\int e^{-\tau (E,z)} \,F(E,z) d\log_{10} (z+1)\;\;\;,
\end{equation}
\begin{equation}
\tau (E,z) = \tau_{\nu \tilde \nu} (E,z) + \tau_{\nu N} (E,z)\;\;.
\end{equation}
Each curve in Fig.\ref{fig:fig3} has a strong cut off on some redshift value. This feature is
connected with the existence in our model the minimum value of PBH mass.
The PBH mass spectrum is steeply falling function of the mass, so the masses near minimum give
a largest contribution to the neutrino background. The moment of their evaporation is, approximately,
\begin{equation}
t_{ev}\approx \frac{\left(M_{PBH}^{min}\right)^3}{3\alpha}\;\;,
\end{equation} %
and the corresponding redshift is determined by the relation 
\begin{equation}
z_{ev}+1\approx (z_{eq}+1)\left(\frac{t_{eq}}{t_{ev}}\right)^{1/2}\;\;\;.
\end{equation}
Larger masses evaporate at larger times and smaller redshifts. If, for instance, $T_{RH}=10^{10}\;\;\mbox{GeV}$,
one has $t_i=0.24\cdot 10^{-26}\;\;\mbox{s}$ ; $M_i=7,5\cdot 10^{10}\;\;\mbox{g}$ , $z_{ev}\sim 10^{7}$.
So, at $T_{RH}=10^{10}\;GeV$ the redshifts  of order of $10^7$ give a largest contribution to the neutrino background
(if there is no neutrino absorption; the absorption effects are essential just near $z\sim 10^7$, and, as one can see from the figure,
the actual $z$-dependences of the integrand have peak values at $z\sim (2\div 3)\cdot 10^6$, depending on a neutrino energy).

The cut off value $z_{ev}$ strongly depends on $T_{RH}$: 
\begin{equation}
\label{414}
z_{ev}\sim\frac{1}{\sqrt{t_{ev}}}\sim\left( M_{BH}^{min}\right)^{-3/2}\sim 
t_i^{-3/2}\sim T_{RH}^{3}\;\;.
\end{equation}
Again, this estimate does not take into account the absorption effects. The real situation is seen on the figure.
In particular, at $T_{RH} > 10^{10}\text{ GeV}$ this cut off is the same for all $T_{RH}$ values, as a result  of a strong
absorption at $z\agt 3\cdot 10^{6}$.

On Figs.\ref{fig:fig4ab},\ref{fig:fig10} spectra of electron neutrino background are shown separately for several channels of the neutrino production.
One can see from Fig.\ref{fig:fig4ab} that 1) the contribution to the summary background from the
direct $\nu_e$ emission is, at $T_{RH} \agt 10^9 \text{GeV}$, absolutely dominant for $E\gtrsim 1 \text{ MeV}$ and 2) the relative
importance of different channels changes with an increase of $T_{RH}$. It is seen also that at $T_{RH} \agt 10^9\text{GeV}$
the contribution of the quark fragmentation channel is very small at $E\gtrsim 1 \text{ MeV}$
$(\lesssim 5 \% )$, so the evident underestimation of this channel in our calculation, connected with the neglect
of the contribution of heavy quark fragmentations, has (at large $T_{RH}$ values) no particular importance.

Situation with the contribution from quark fragmentations is slightly different at smaller $T_{RH}$ values
(see Fig.\ref{fig:fig10}). Due to the flattening out of the curve for direct $\nu_e$ emission the
relative contribution of the fragmentation channel at $E\sim 1\text{MeV}$ grows with an decrease of $T_{RH}$. The highest 
relative contribution is at $T_{RH} \sim 10^{8.5}\text{GeV}$. At this $T_{RH}$ the  minimum value of mass in PBH
mass spectrum ($M_{BH}^{min}$) coincides with the threshold of quark evaporations ($m_{thr}\sim 10^{14.1}\text{g}$, 
according to Fig.\ref{fig:fig1}). At further decrease of $T_{RH}$ the relative contribution of quark fragmentations
at $E\sim 1 \text{MeV}$ fastly decreases and, at some $T_{RH}$ value, it is strictly zero (when, due to correspondingly
large $M_{BH}^{min}$ value, the age of the universe is not enough for "warming-up" the PBHs). 

It follows from Fig.\ref{fig:fig10} that at neutrino energies larger than $\sim (5\div 10)\text{MeV}$ the
contribution of fragmentation channels is negligibly small ( $\lesssim 5\%$). As was mentioned above, the region
of smaller neutrino energies is not  important for the calculation of spectral index constraints, so some
possible underestimation of the quark fragmentation contribution in this region (due to the simplified treatment
of them in the present paper) is not sufficient.

Some typical results of calculations of summary electron neutrino background spectra are shown on 
Figs.\ref{fig:fig5},\ref{fig:fig6}.
The main features of such spectra have been revealed in previous works \cite{12,26,27}: $E^{-3}$
dependence above  $100 \text{ MeV}$ and the flattening out of the spectra at lower energies.
The form of background spectra depends on the form of an initial PBH mass distribution.
In refs.\cite{26,27} the simple power mass spectrum of PBHs  containing no cut off (minimum mass value)
were used. The behavior of background spectra at low energies in this case is, aside from jet 
fragmentation contribution, $\sim E^{-1}$ and turnover energy is $\sim 100\text{MeV}$ \cite{26}.
In the case of PBH mass spectrum with the cut off the turnover energy depends on $T_{RH}$ \cite{12}
(it decreases as $T_{RH}$ increases, as is seen at Fig.\ref{fig:fig5}). According to Figs.\ref{fig:fig5},\ref{fig:fig6} the 
turnover energy is about $1\text{ MeV}$ at $T_{RH}=10^9 \; \text{ GeV} $.
Besides, it was shown in ref.\cite{12} that at low energies the photon background spectra are almost flat
and,as is seen in Figs.\ref{fig:fig5},\ref{fig:fig6}, the same is also true in the case of the neutrino
background spectra.

\section{Constraints on the spectral index}
\label{sec:constr}
For an obtaining of the constraints on the spectral index we use three types
of neutrino experiments.

1. {\it Radiochemical experiments for the detection of solar 
neutrinos.} There are data on solar neutrino fluxes from the famous Davis experiment \cite{20} and the
$Ga-Ge$ experiment \cite{21}. The basic process used in radiochemical experiments is a 
charged-current neutrino capture reaction
\begin{equation}
\nu_e + A^N_Z \to e^- + A^{N-1}_{Z+1} \; .
\end{equation}

In general, the cross section for the neutrino absorption via
a bound-bound transition can be calculated using the approximate formula
\begin{equation}
\label{qwert}
\sigma = \left\{
      \begin{aligned}[l]
      &\frac{G_F^2}{\pi}\left(\langle 1 \rangle^2+\left(\frac{g_a}{g_v}\right)^2
      \langle\sigma\rangle^2\right)p_e E_e &\; , \;& E<100MeV\\
      &\text{\it const. } &\; , \;& E > 100MeV .
      \end{aligned}\right.
\end{equation}
Here, $p_e$ and $E_e$ are the momentum and the energy of an electron produced  in a neutrino
capture reaction.  Neutrino and electron energies are connected by the relation
\begin{equation}
E_e=E-E_{thr}+m_e\;.
\end{equation}
Stopping of the rise of a neutrino capture cross section (for a bound-bound transition)
with neutrino energy is due to influence of a weak nuclear formfactor of the transition.
The approximate value of the turnover energy in Eq.(\ref{qwert}) is determined from the relations
\begin{equation}
E\sim q_{max}\sim \frac{1}{R_A} \sim m_{\pi}\;.
\end{equation}  
Here, $q_{max}$ is a maximum momentum transfer from a neutrino to the nucleus
(in a model with step-like nuclear formfactor), 
$R_A$ is a size of the nucleus.

In the case of the Davis experiment the neutrino capture reaction is
\begin{equation}
\nu + ^{37} Cl \to ^{37}Ar + e^{-} \; .
\end{equation}
Here we took into account the super- allowed transition only, i.e., the transition 
to the isotopic analog state in $^{37} Ar$, for which   
\begin{equation}
\langle 1 \rangle^2=3\;\; ,\;\; \langle\sigma\rangle^2=0.2\;\; ,\;\; E_{thr}\approx 5MeV.
\end{equation} 
This transition gives dominant contribution to the effect in the detector, in spite of the
rather high energy threshold, due to the large value of the Fermi strength $\langle 1 \rangle^2$.

In the case of the $Ga-Ge$ reaction, 
\begin{equation}
\label{swe}
\nu + ^{71}Ga \to ^{71}Ge +e^{-}\;,
\end{equation} 
the main contribution 
gives the ground state - ground state
transition ($E_{thr}=0.242MeV$). In $^{71}Ge$ also there is an isotopic analog state, but its 
energy is too high, $\sim 9\text{ MeV}$ (it is slightly higher than the particle emission threshold, so this 
transition is irrelevant).
The cross section of the ground state - ground state transition in the reaction (\ref{swe})
is given by the expression \cite{42} (for $E< 100 \text{ MeV}$)  
\begin{equation}
\sigma \cong \frac{G^2}{\pi} p_e E_e \frac{2 J_A+1}{2J_B + 1}\cdot \frac{6163.4}{f t}\;.
\end{equation}
Here, $J_A$ and $J_B$ are angular momenta of $^{71}Ga$ and $^{71}Ge$, respectively, $f t$
value is determined by the half-life time of $^{71}Ge$. According to ref.\cite{42} one has 
\begin{equation}
J_A=\frac{3}{2}\;\;,\;\;J_B=\frac{1}{2}\;\;,\;\;\lg ft = 4.37\;\;.
\end{equation}  

The number of target atoms $N^T$ in the detectors is about $2.2\cdot 10^{30}$ for the
$Cl-Ar$ experiment \cite{20} and $\sim 10^{29}$ for the $Ga-Ge$ experiments \cite{21}.
The average statistics is $\sim 1.5\; \text{ captures/day}$ ($Cl-Ar$) and $\sim 1\; 
\text{ capture/day} $ ($Ga-Ge$). The constraints are calculated using the relation
\begin{equation}
4\pi\cdot N^T\cdot10^5\cdot\int S(E)\sigma(E)dE<1.
\end{equation}

2. {\it The experiment on a search of an antineutrino flux from the Sun} \cite{22}.
In some theoretical schemes (e.g., in the model of a spin - flavor precession in
a magnetic field) the Sun can emit rather large flux of antineutrinos. LSD
experiment \cite{22} sets the upper limit on this flux, $\Phi_{\tilde \nu}/\Phi_{\nu}
\le 1.7\%$. In this experiment the neutrino detection is carried out using
the reaction
\begin{equation}
\label{reaction}
\tilde \nu _e + p \to n + e^+
 \; .
\end{equation}
The number of target protons is $\sim 8.6\cdot 10^{28}$ per 1 ton of the
scintillation detector,  and the obtained upper limit is $0.28$ antineutrino
events per year per ton \cite{22}. The cross section of the reaction~(\ref{reaction})
is well known (see, e.g.,\cite{23}). In particular, in the low energy region ($E\ll m_p$) it is
given by the formula
\begin{equation}
\sigma(\tilde \nu + p + e^{+} + n)=\frac{G^2_F}{\pi} \left(1+3\left(\frac{g_a}{g_v}\right)^2 
\right)p_e E_e,
\end{equation}
\begin{equation}
E_e=E-(m_n-m_p).
\end{equation} 
The threshold of this reaction is given by the relation
\begin{equation}
E_{thr}= m_n-m_p+m_e\approx 1.8\mbox{ MeV}.
\end{equation}
The cross section of the reaction (\ref{reaction}) grows 
with the neutrino energy up to $E\sim 2\;GeV$, and is a 
constant ($\sim 0.5\cdot 10^{-38}\;cm^{2}$)
at larger energies.The product of this cross section 
and a PBH antineutrino background spectrum has the 
broad maximum at $E\sim 100MeV$. The constraint is determined 
from the condition that the calculated effect in the {\bf LSD}
detector is smaller than the upper limit obtained in ref.\cite{22}.

3. {\it The Kamiokande experiment on a detection of atmospheric electron 
neutrinos }\cite{24}. In this experiment the electrons arising in the reaction  
\begin{equation}
\nu_e^{atm}+n\to p+e^-
\end{equation}
in the large water Cherenkov detector were detected and, moreover, their 
energy spectrum was measured. This spectrum has a maximum at the energy
about $300MeV$. The spectrum of the atmospheric electron neutrinos is
calculated(see, e.g.,\cite{25}) with a very large accuracy (assuming an absence of the neutrino 
oscillations) and the experimentally measured electron spectrum coincides,
more or less, with the theoretical prediction. The observed electron excess
at $E\sim 100MeV$ (which is a possible consequence of the oscillations) 
is not too large. We use the following condition for an obtaining  the our 
constraint : the absolute differential intensity of the PBH neutrino 
background at the neutrino energy $E\sim 0.3GeV$ cannot exceed the theoretical
differential intensity of the atmospheric electron neutrinos at the same 
energy (otherwise the total number of electrons in the detector and the energy 
spectrum of these electrons must be different from the observed ones).

\section{Results and discussions}

\label{sec:res}

Fig.\ref{fig:fig7},\ref{fig:fig9} shows our results for the spectral index constraints. It is seen that 
the best constraints are obtained using the Kamiokande atmospheric neutrino data and
the LSD upper limit on an antineutrino flux from the Sun. 

It appears that the spectral index constraints obtained  in the  present work are not  stronger than
the corresponding constraint following from COBE data  (see Eq.(\ref{g001})). One  should remember,
however, that our approach directly constrains the perturbation amplitude $\sigma_H (M_h)$ rather than
the spectral index. Comoving  length scales of perturbations, for which we constrain the amplitude, 
are very small  in  comparison  with the size of present horizon. For example, if $T_{RH}\sim  3\cdot 10^8 \text{ GeV}$
the minimum PBH mass value is about $10^{14}\text{ g}$. In this case  the time of PBH formation
is 
\begin{equation}
t_i\sim  \frac{M_{BH}^{min}}{M_{pl}} t_{pl}\sim 10^{-24} \text{ s},
\end{equation} 
the corresponding scale factor  is, according  to Eq.(\ref{sdw}), about$\sim 10^{22}$, and  
for the comoving  length scale one obtains
\begin{equation}
\lambda_i\sim \frac{c t_i}{a}\sim 10^{-9} \text{ pc}.
\end{equation}
It is clear that the normalization on COBE data has sense only if  it is assumed that the spectral index
is independent  on scale. The   curves of  Fig.\ref{fig:fig7},\ref{fig:fig9} were obtained  using
just this assumption; it  is very easy to  calculate from  them  the corresponding constraints
on $\sigma_H (M_h)$.

For a demonstration of the sensitivity of these results to a chosen $\Omega_{\Lambda}$-value  we calculate
the constraints for two extreme $\Omega_{\Lambda}$-values, for which the parametrization for 
$f_S^{(0)}$ (Eq.(\ref{d16b})) is still valid. It is clear that some weakening of the constraint
with an increase of $\Omega_{\Lambda}$ is due to the corresponding decrease of values of the 
function $\sigma_{H} (M_h)$ entering the initial PBH mass spectrum (see Eq.(\ref{220})). This 
decrease of $\sigma_H (M_h)$ takes place solely due to the dependence of
$f_S^{(0)}$ on $\Omega_{\Lambda}$ (see Eq.(\ref{d16b})).

The behavior of the 
constraint curve $n(T_{RH})$ is sharply different from that was obtained 
in ref. \cite{13} (authors of ref. \cite{13} used the initial PBH mass function following
from the near critical collapse scenario with a domination of the earliest 
epoch of PBHs formation, and diffuse extragalactic photon data).

The slight bend at $T_{RH}\sim 10^{10}\text{GeV}$ of the constraint curves
(Fig.\ref{fig:fig7}) is an effect of a neutrino absorption in the space.
Constraint based on the atmospheric neutrino  experiment  is  especially sensitive to
the  absorption because neutrinos of larger energies  (in average) are responsible for it. 

Flattening out of the constraint curve $n(T_{RH})$ with increase of $T_{RH}$ is  
connected with the dependence $M_{BH}^{min}(T_{RH})$. According to the relations
(\ref{c6},\ref{g622},\ref{l1}), the minimum  value of PBH mass in the initial mass spectrum
is inversely proportional to $T_{RH}^2$.  More exactly, the relation between $M_{BH}^{min}$ and $T_{RH}$ is
\begin{equation}
M_{BH}^{min}(g)=\frac{7.2\cdot 10^{30}}{(T_{RH}/1 GeV)^2}.
\end{equation}
 At small $T_{RH}$ the minimum
value of PBH mass in the initial PBH mass spectrum is large and the neutrino background
is dominated by the evaporations at recent epochs. In opposite, at large $T_{RH}$, when
$M_{BH}^{min}$ is small, the neutrino background is dominated by the evaporations at earlier times.
Thus, using the relation
\begin{equation}
z_{eq}\simeq 2.4 \cdot 10^{4} \Omega_m h^2 
\end{equation}
and the curves, corresponding to $T_{RH}=10^9\text{GeV}$ in Fig.\ref{fig:fig3}, one
can see that in cases when $T_{RH}>10^9\text{GeV}$ the neutrino background is dominated by
PBH evaporations during the radiation era.

Introducing, as usual \cite{51}, the function $M^* (M_{BH}^{min})$ by the formula
\begin{equation}
M^* (M_{BH}^{min})=\left[ 3\alpha (M_{BH}^{min}) t_0 \right]^{1/3} \; ,
\end{equation}
where $t_0$ is the age of the universe, one can easily see that at low $T_{RH}$
(before the flattening out of our $n(T_{RH})$ curve) the inequality $M_{BH}^{min}> M^*(M_{BH}^{min})$
takes place while at large $T_{RH}$ one has $M_{BH}^{min}< M^* (M_{BH}^{min})$. We remind that $M^*$
is the mass of a black hole, formed in the early universe and evaporating today. The 
turning-point (i.e., the point of a change of  the slope  of the $n(T_{RH})$ curve)
corresponds to the equality $M_{BH}^{min}=M^{*}(M_{BH}^{min})$ and 
 in the case $\Omega_{\Lambda} =0$ it is at $T_{RH}\sim 10^8\text{GeV}$.
The shift of this turning-point with an increase of $\Omega_{\Lambda}$ is explained by 
the corresponding increase of an age of the universe:
\begin{equation}
t_0=\frac{2}{3 H_0 \sqrt{\Omega_{\Lambda}}}\ln\frac{1+\sqrt{\Omega_{\Lambda}}}{\sqrt{1-\Omega_{\Lambda}}}
\; \stackrel{\Omega_{\Lambda}\to 0}{\longrightarrow} \; \frac{2}{3 H_0} \; .
\end{equation}

One can compare our spectral index constraints with the corresponding results
of ref. \cite{12}, where the same initial mass spectrum of PBHs had been used. Some
difference in resulting constraints ($\Delta n\sim 0.02$) may be connected simply 
with the fact that authors of ref. \cite{12} use slightly different formula for
$\sigma_H (M_h)$, namely,
\begin{equation}
\sigma_{H}(M_h)=\sigma_H(M_{h\;,0})\left(\frac{M_h}{M_{h\;,0}}\right)^{\frac
{1-n}{4}}.
\end{equation}
In other respects the constraints are quite similar although in ref. \cite{12} 
they were based on diffuse extragalactic photon background data. 

One should note that 
usually the calculations of these constraints are accompanied by the calculation
of the bounds based on requirement that the energy density in PBHs does not overclose
the universe at any epoch ($\Omega_{BH} < 1$). For a setting of such bounds one 
must consider the cosmological evolution of the system PBHs + radiation. We 
intend to carry out these calculations in a separate paper. Estimates show that 
at large $T_{RH}$ values ($T_{RH} \agt 10^{10}\text{GeV}$) the constraints based on
$\Omega_{BH} <  1$ are stronger than those based on the neutrino experiments (it is the reason
why we did not explore the region $T_{RH} > 10^{10}\text{GeV}$ in the present paper).

\begin{figure}[!t]
\epsfig{file=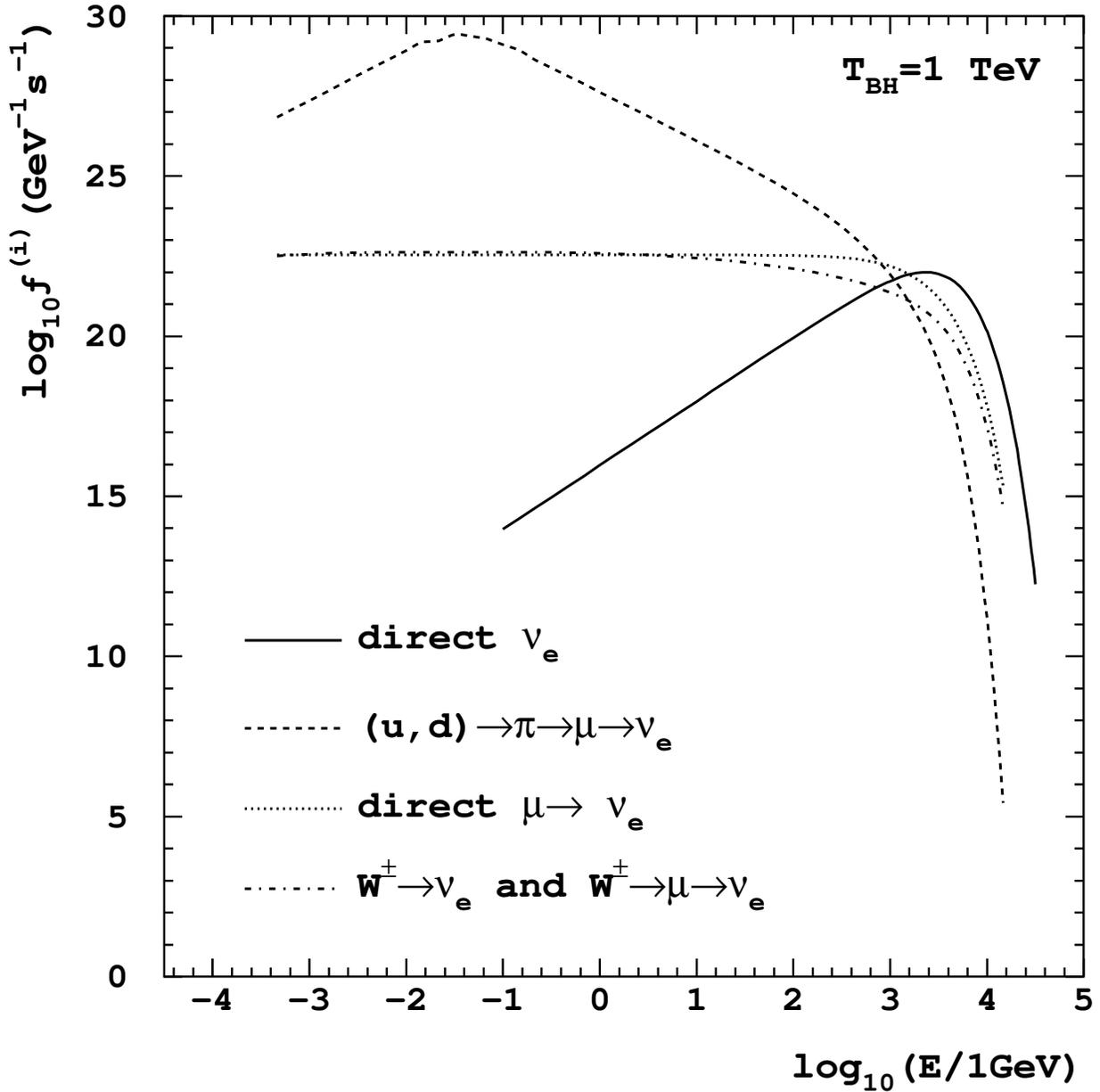,width=\columnwidth}
\caption{Instantaneous electron neutrino spectra from the evaporation of PBH 
with $T_{BH}=1\text{TeV}$, 
for several channels of the neutrino production.}
\label{fig:fig2}
\end{figure}

\begin{figure}[!t]
\epsfig{file=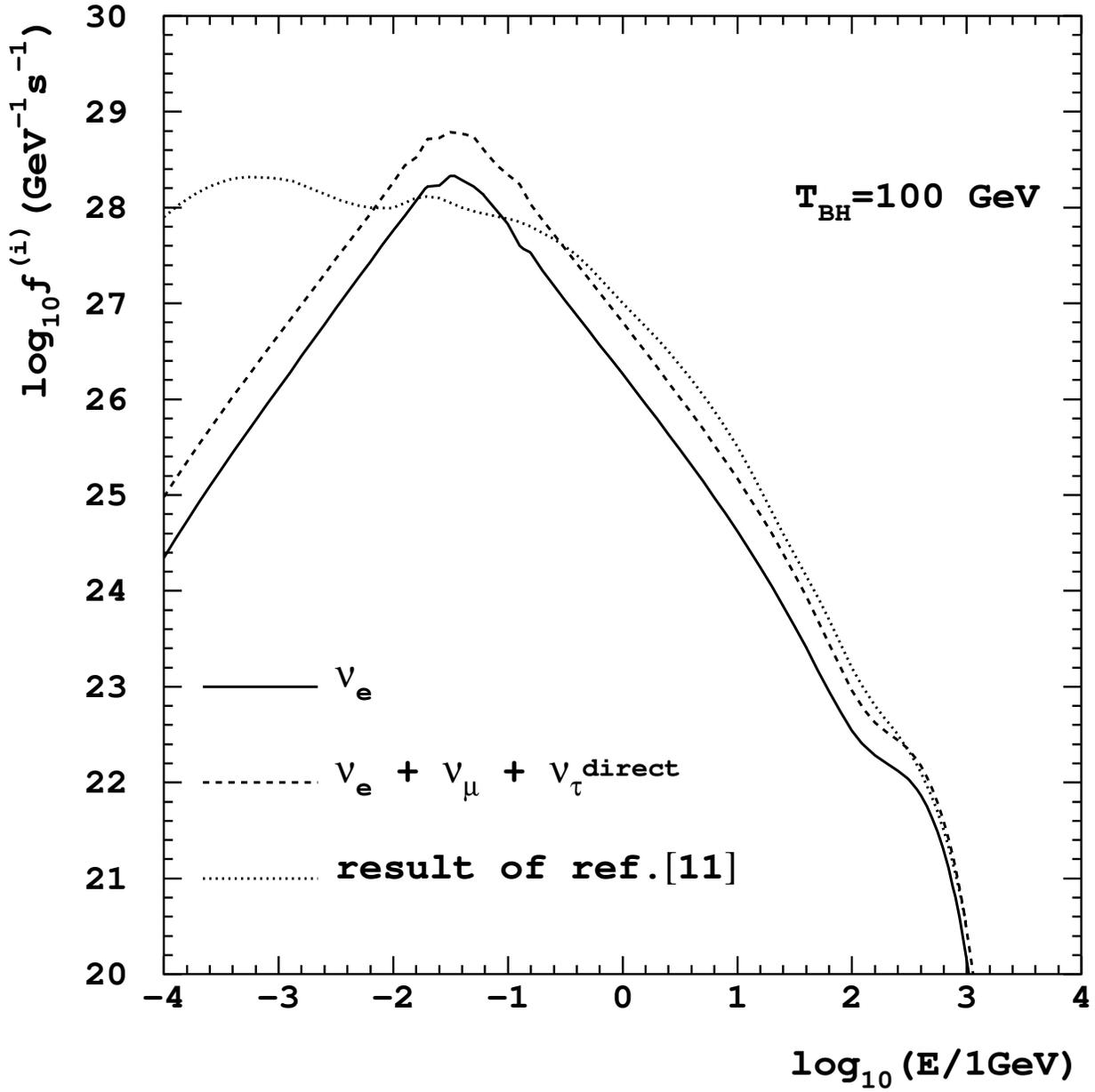,width=\columnwidth}
\caption{ Instantaneous neutrino spectra  from a black hole with a  temperature of 100 GeV.}
\label{fig:fig11}
\end{figure}

\begin{figure}[!t]
\epsfig{file=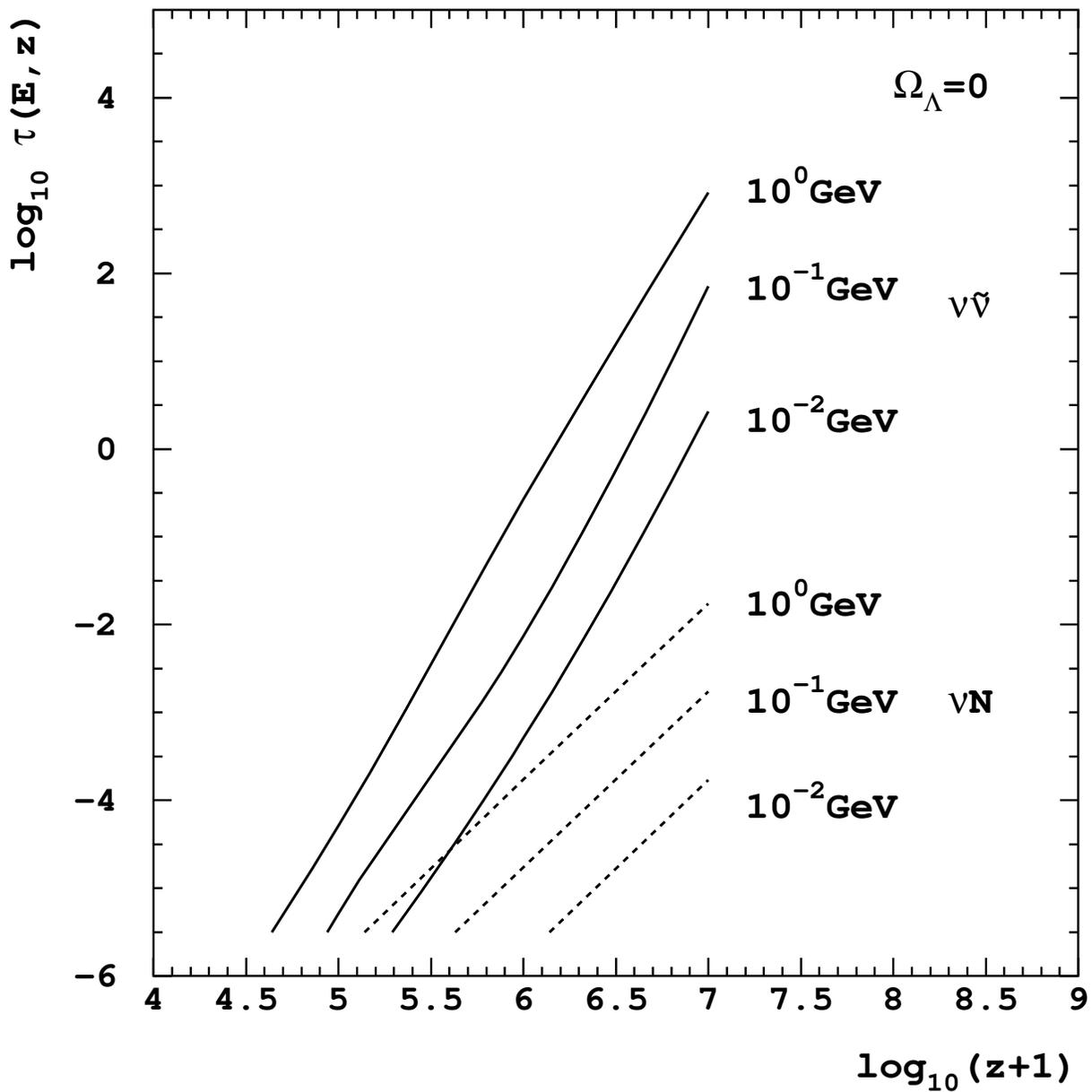,width=\columnwidth}
\caption{Absorption factor for two channels of neutrino absorption and for several 
values of neutrino energy, as a function of $z$, the redshift corresponding to a 
moment of the neutrino emission.}
\label{fig:fig8}
\end{figure}

\begin{figure}[!t]
\epsfig{file=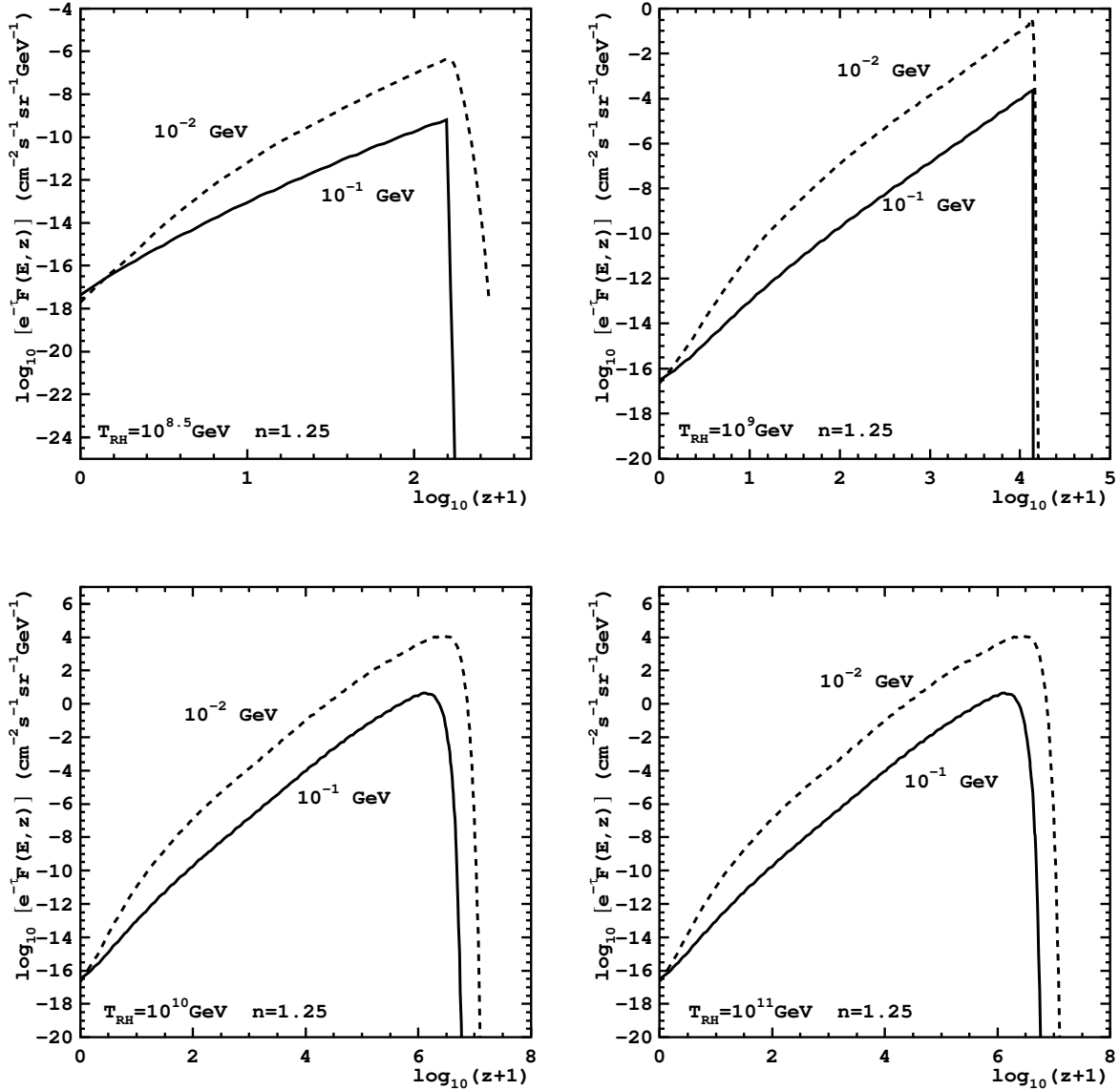,width=\columnwidth}
\caption{Redshift dependence of the integrand of the expression (\ref{sdfr}) for
a neutrino background spectrum (with $\Omega_{\Lambda}=0$), for two values of the neutrino energy and for several 
values of the parameter $T_{RH}$ .}
\label{fig:fig3}
\end{figure}

\begin{figure}[!t]
\epsfig{file=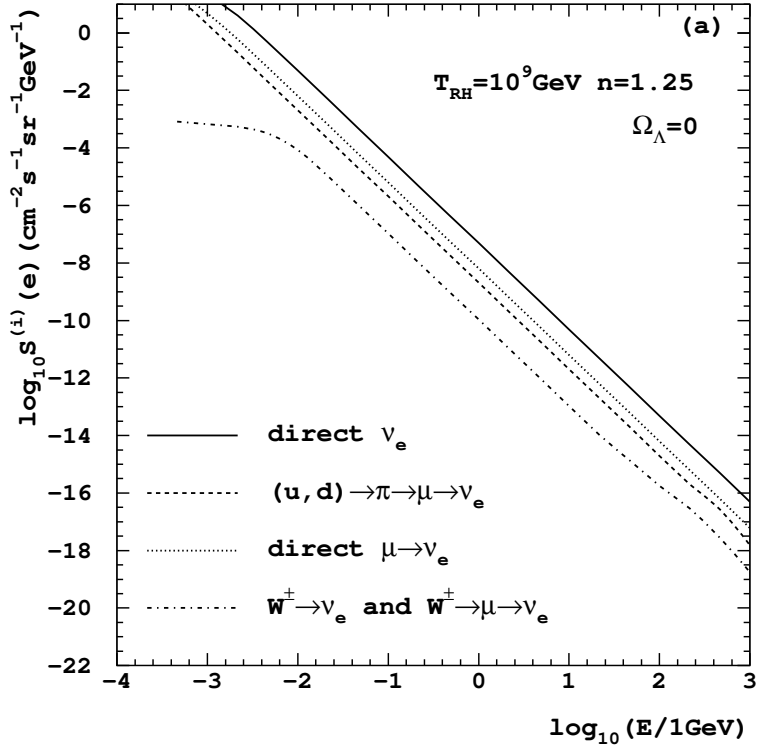,height=11cm}
\epsfig{file=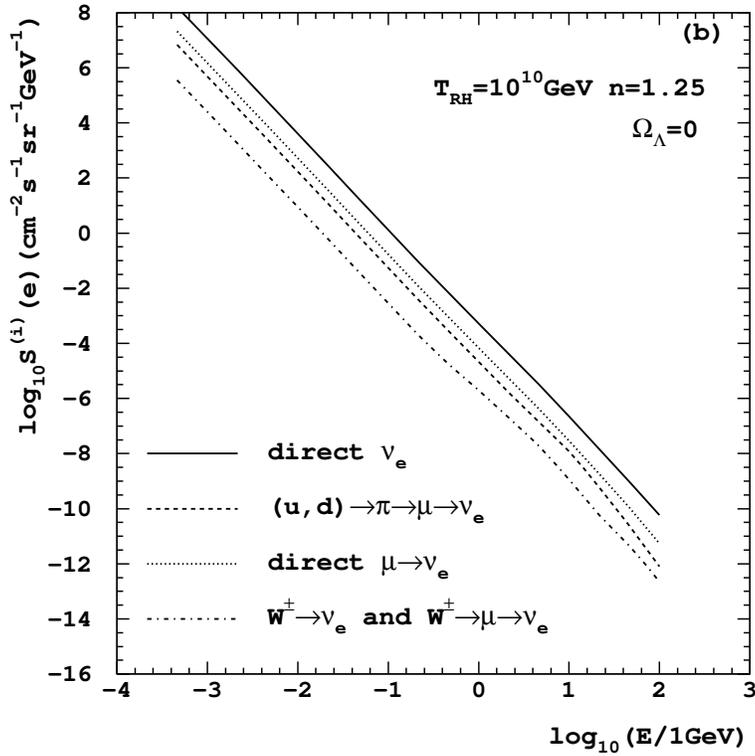,height=11cm}
\caption{Separate contributions to neutrino background spectra 
from different channels of the neutrino production during PBH 
evaporation, for fixed values of the parameters $T_{RH}$ and $n$.}
\label{fig:fig4ab}
\end{figure}

The neutrino background spectra obtained in the present paper may be useful somewhere else, irrespective
of the spectral index constraint problem. The main feature of such spectra at $E\agt 1 MeV$ and at $T_{RH}\agt 10^9-10^{10}\text{ GeV}$
is the very small contribution ($\alt 5 \%$) of all secondary channels of the neutrino production at evaporation 
(except the decay of directly evaporated muons) (see Fig.\ref{fig:fig4ab}).
There are two reasons for this: the steepness of the initial PBH mass spectrum and, most essentially,
the existence in our model the cut off in this spectrum, i.e., the minimum PBH mass value.
Evaporation of PBHs having masses near $M_{BH}^{min}$ gives the dominant contribution in the 
background, so there is a strong maximum in $z$ distributions at low energies (Fig.\ref{fig:fig3}). 
For example, at $T_{RH}=10^{10}\text{GeV}$ the peak value
of $z$ is $\sim 10^6$. Neutrino energy $\sim 1\text{MeV}$ corresponds  
in this case to energy 
at evaporation $\sim 1\text{TeV}$. For an evaporation of such high energy neutrinos the PBH mass 
must be $\sim 10^{10}\text{g}$ in the case of direct neutrinos and much smaller in the case of secondary
neutrinos (because their energies are much lower in average (see Fig.\ref{fig:fig2})). At the same time, 
the minimum value of PBH mass
in the mass spectrum at $T_{RH}=10^{10}\text{GeV}$ is $\sim 10^{11}\text{g}$. So, it should be clear
that, in our example, the secondary neutrinos with $E\sim 1\text{MeV}$ can appear only after
the essential evolution of the initial PBH mass spectrum (see Eq.(\ref{311}) and the detailed discussion
of this evolution in ref.\cite{50}). This evolution leads to an increase of the total mass interval 
and to the corresponding decrease of a PBH number density (per unit mass interval) and, as a result,
to a relative smallness of the secondary neutrino flux.

\begin{figure}[!t]
\epsfig{file=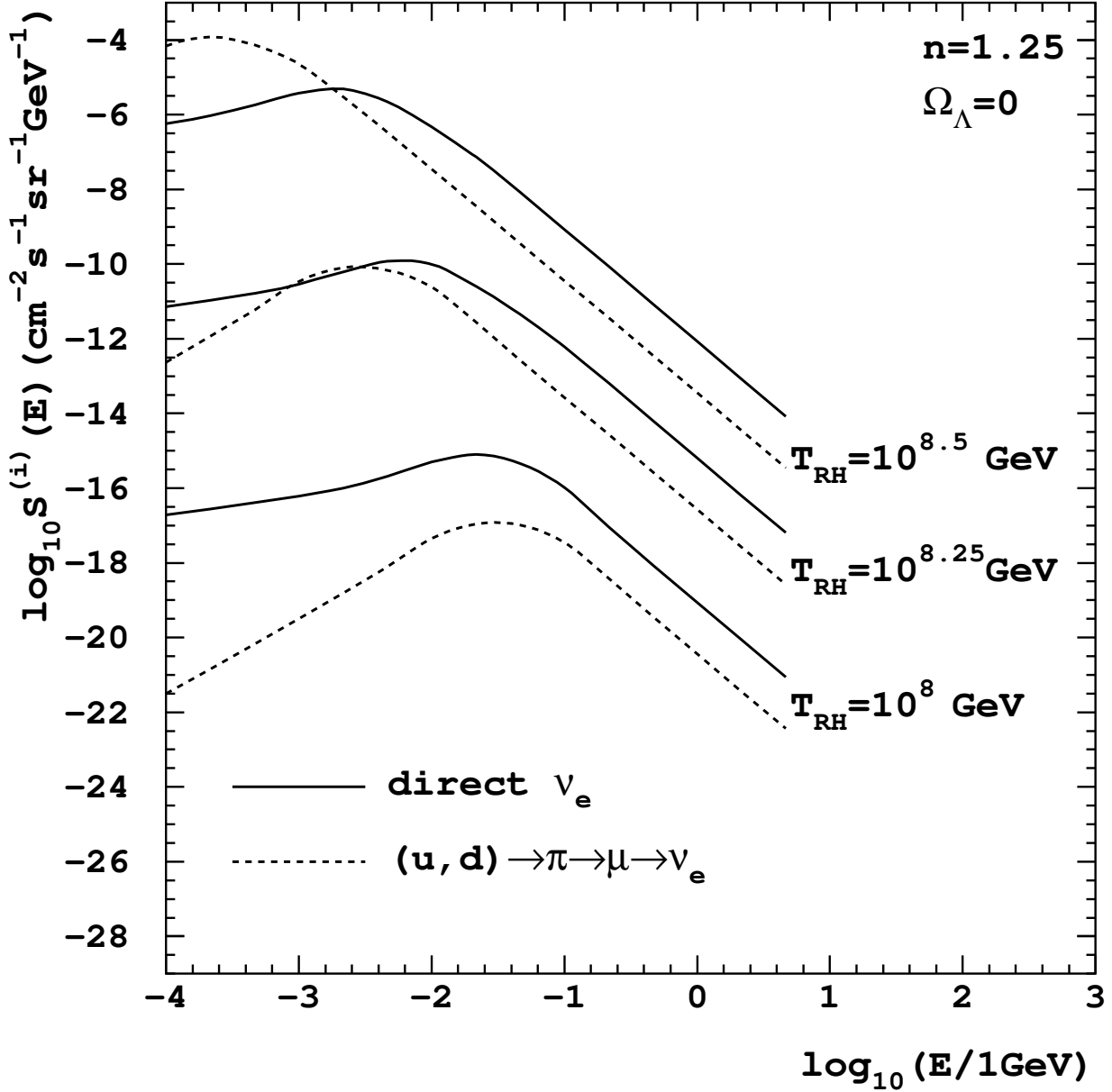,width=\columnwidth}
\caption{Separate contributions to neutrino background spectra from two channels of the neutrino
production during PBH evaporation, for three different values of $T_{RH}$.}
\label{fig:fig10}
\end{figure}

\begin{figure}[!t]
\epsfig{file=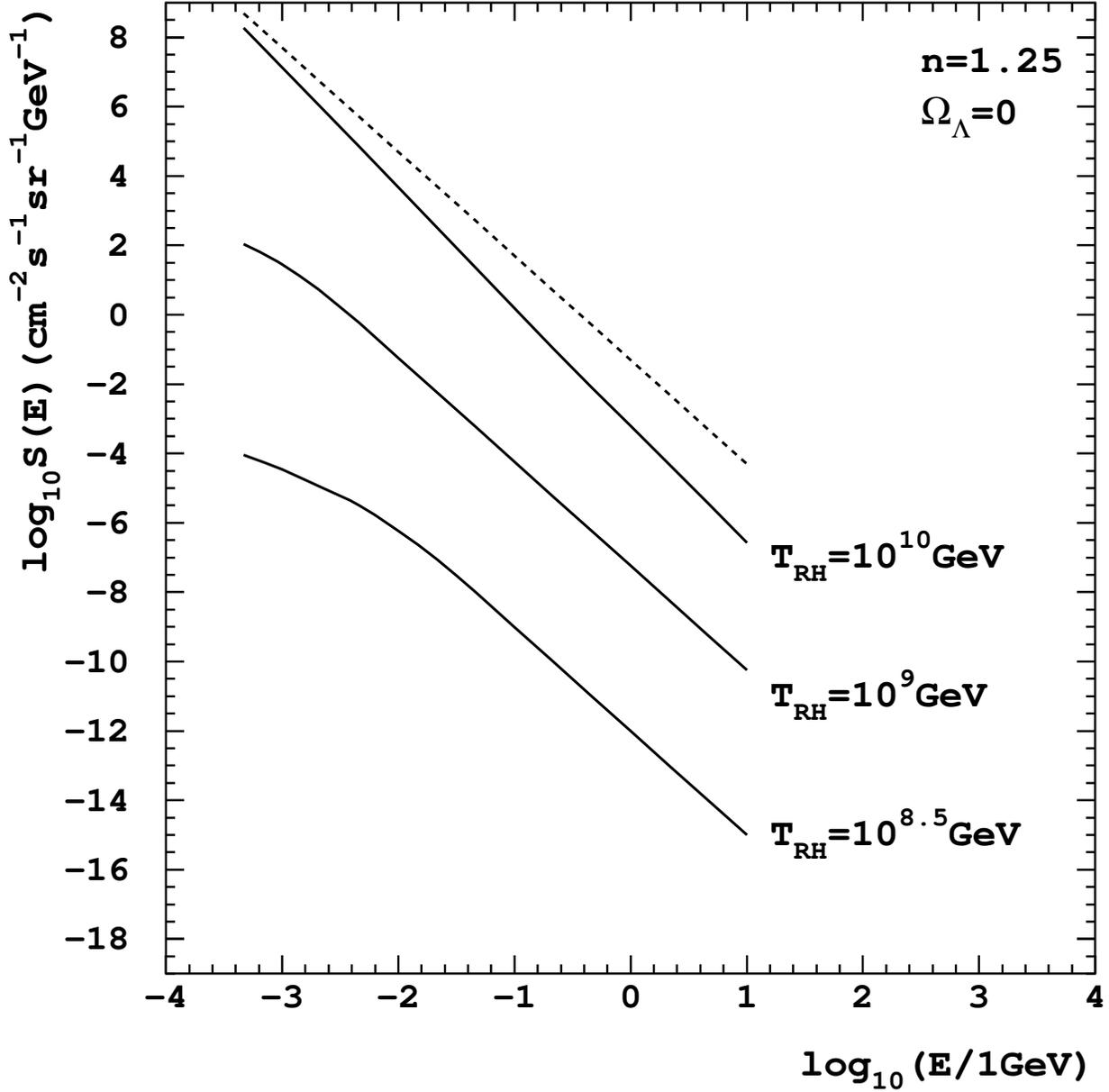,width=\columnwidth}
\caption{Electron neutrino background spectra from PBHs,
calculated for several values of reheating temperature. 
Dashed curve shows the spectrum for $T_{RH}= 10^{10}\text{GeV}$, 
calculated without taking into account the neutrino absorption.} 
\label{fig:fig5}
\end{figure}

\begin{figure}[!t]
\epsfig{file=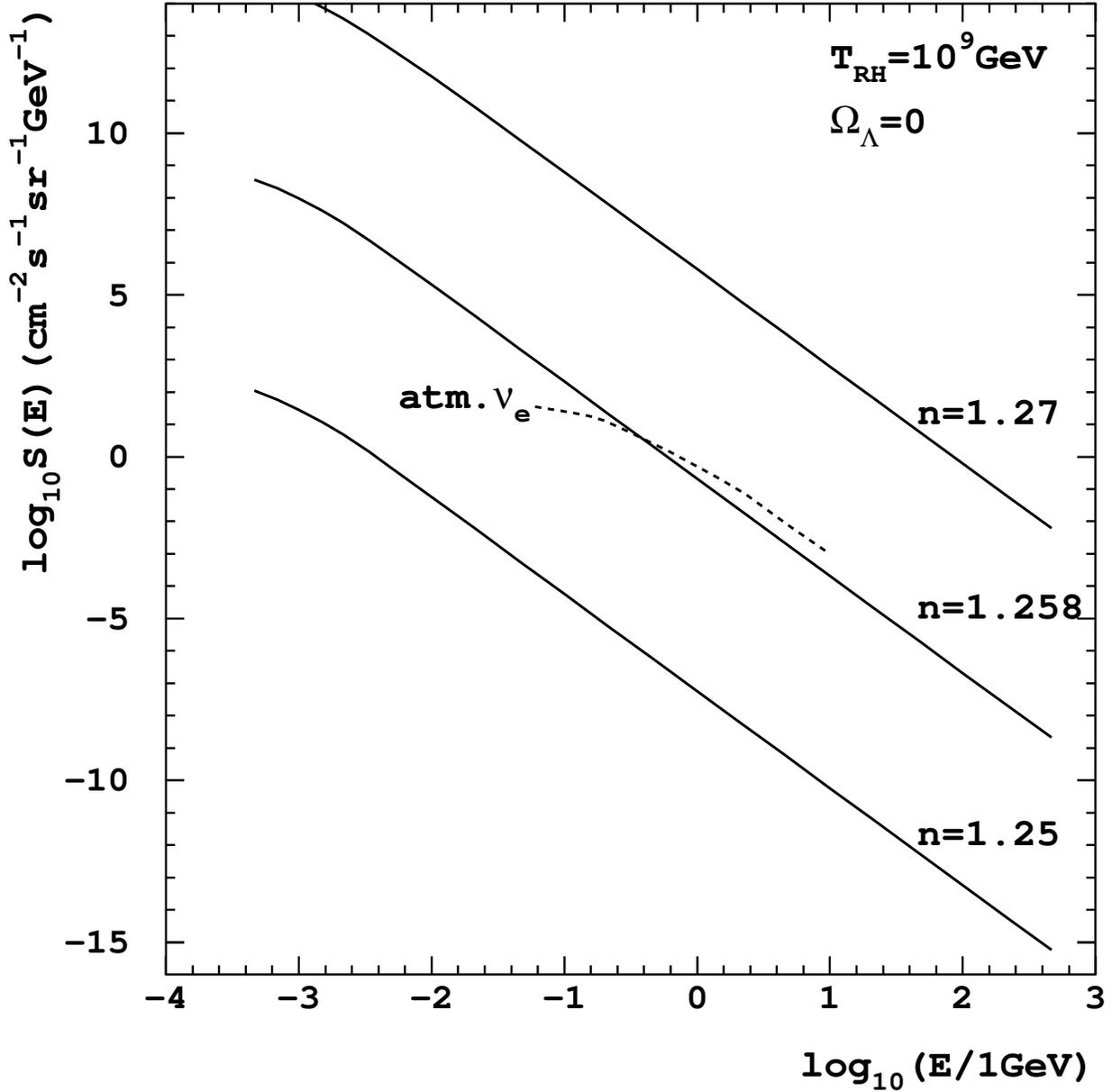,width=\columnwidth}
\caption{Electron neutrino background spectra from PBHs,
calculated for several values of the spectral index.
Dashed curve shows the theoretical atmospheric neutrino
spectrum at Kamiokande site~[39] (averaged over all directions).}
\label{fig:fig6}
\end{figure}

\begin{figure}[!t]
\epsfig{file=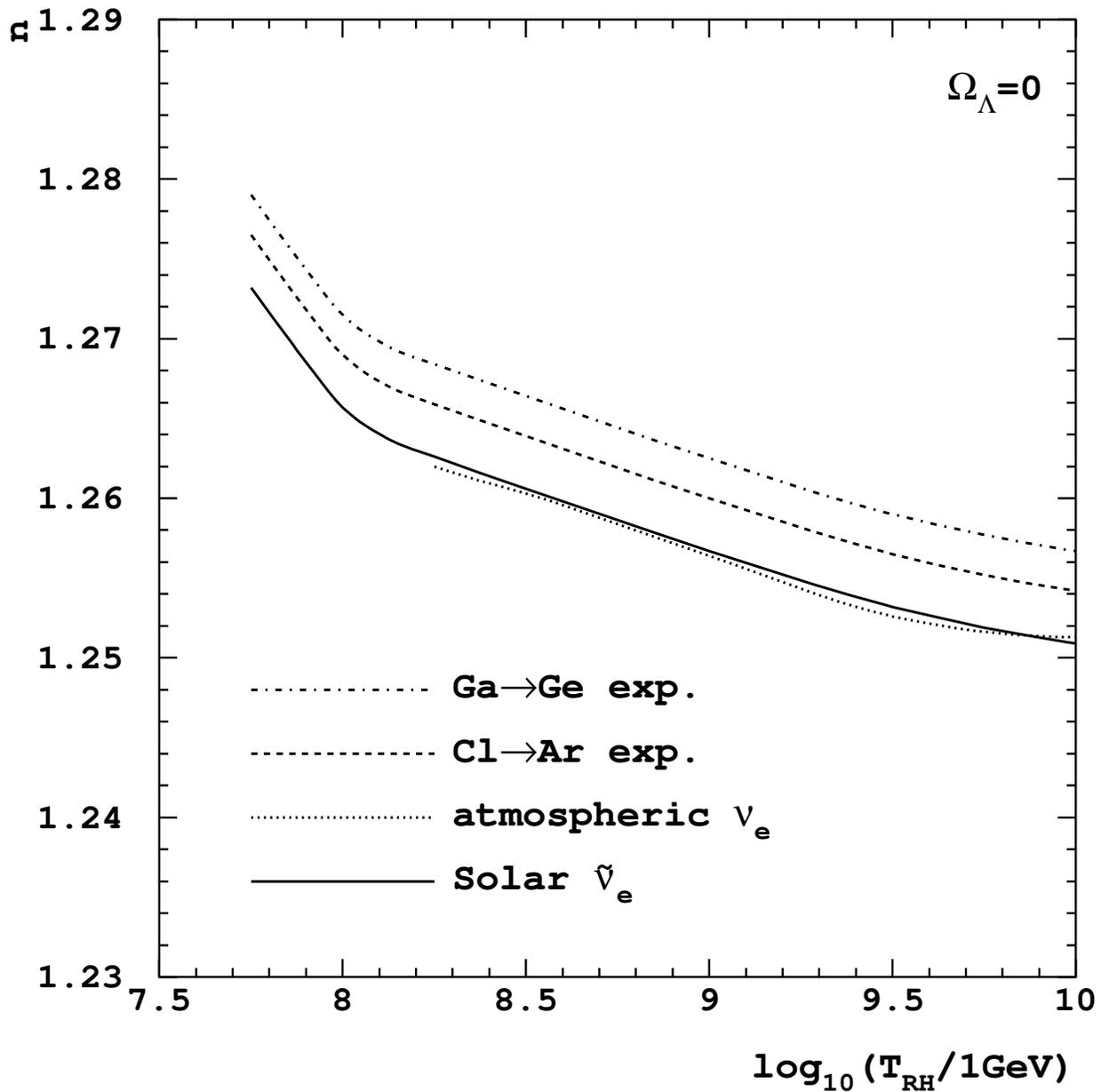,width=\columnwidth}
\caption{Constraints on the spectral index $n$ as a function of the 
reheating temperature $T_{RH}$, from three types of the neutrino experiments.}
\label{fig:fig7}
\end{figure}

\begin{figure}[!t]
\epsfig{file=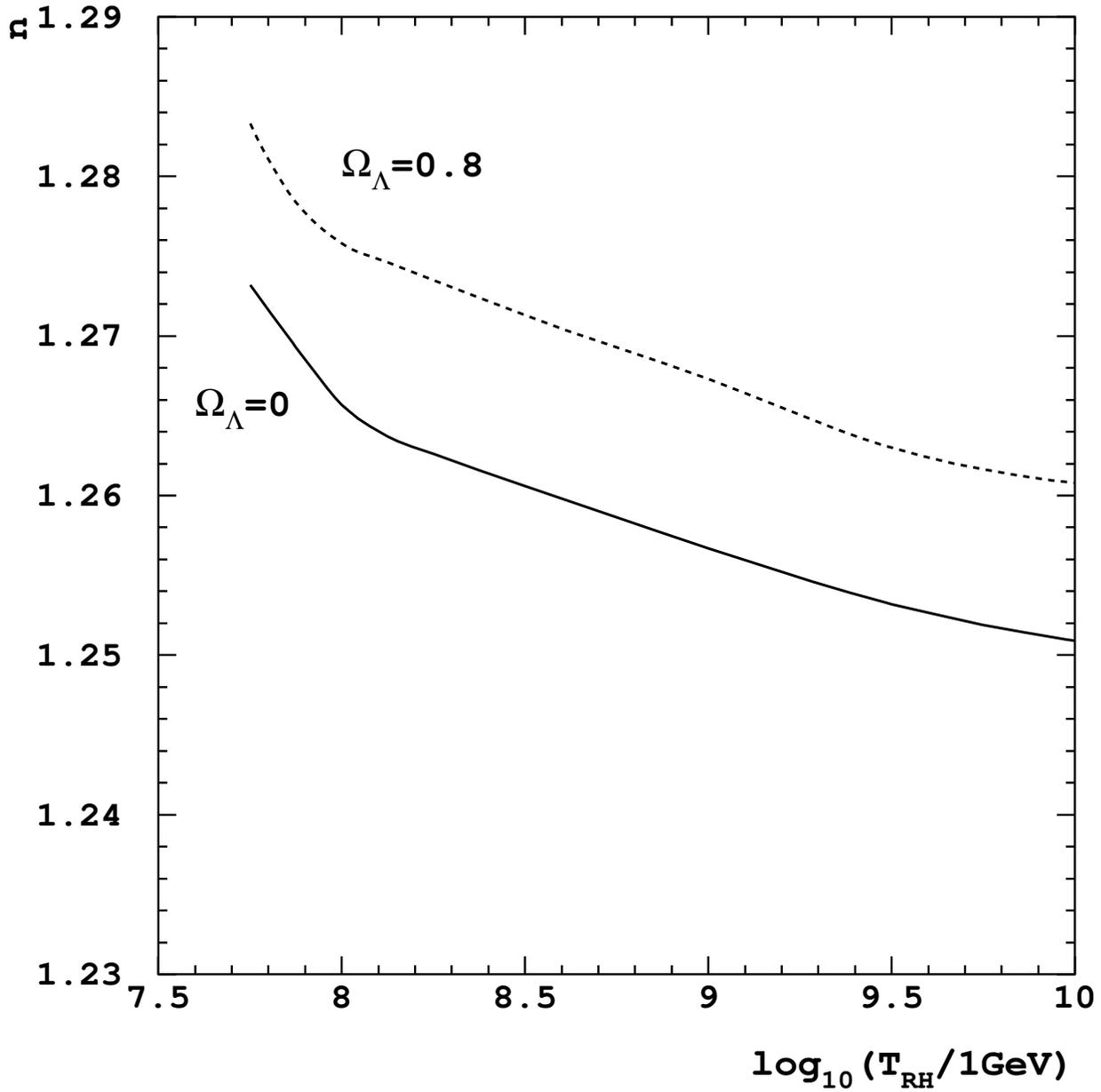,width=\columnwidth}
\caption{Constraints on the spectral index $n$ as a function of the 
reheating temperature $T_{RH}$, based on the solar $\tilde \nu_e$ experiment, for two values
of $\Omega_{\Lambda}$.}
\label{fig:fig9}
\end{figure}

One can compare our results for the neutrino background spectra with those of ref.\cite{26,27}.
According to the  conclusions of ref.\cite{27}, the secondary neutrino fluxes are
larger relatively and even dominant at $E\alt 10\text{MeV}$. Similar results were obtained
in earlier work \cite{26}. In our  model the dominance of secondary neutrino fluxes at $E  < 10 \text{ MeV}$
takes place at some  values of $T_{RH}$ only (see Fig.\ref{fig:fig10}).

The results for neutrino diffuse background spectra and the constraints on the spectral index obtained here
do not take into account the possible existence of the photospheres \cite{28,29,43} around a black
hole. In studies of photon background spectra two kinds of photospheres should be considered: {\bf QCD}
photosphere (quark-gluon plasma arising from interactions of quarks and gluons before hadronization) and
{\bf QED} photosphere ($e^+ -e^--\gamma$ plasma, in which directly evaporated photons are processed).
As for the latter, in our case one should consider instead the "electroweak (or neutrino) photosphere"~\cite{28}
which may form only at extremely high black hole temperatures and, probably, has no influence on the 
background spectra. {\bf QCD} photosphere may be more important for the problem. Thus, the recent
calculations \cite{43} of instantaneous photon spectra from the model {\bf QCD} photosphere show that some
increase of photon fluxes (in comparison with the usual case when {\bf QCD} photosphere
does not form and photons appear simply through the direct quark fragmentations with subsequent $\pi^0$-decays) 
definitely takes place. For example, according to these calculations, at $T_{BH}=50\text{GeV}$ the peak value of instantaneous photon
spectrum increases on a factor $\sim 4$. If it is true and if this increase is the same at all black hole temperatures,
the contribution to neutrino background spectra from the quark fragmentations obtained in our 
calculations must be increased approximately on the same factor.

\acknowledgments

We wish to thank G. V. Domogatsky for valuable discussions and comments. We are grateful also 
to H.I.Kim for informing us about his work on the same problem and for useful remarks.

\newpage

\end{document}